# Examining the Effect of COVID-19 on Foreign Exchange Rate and Stock Market – An Applied Insight into the Variable Effects of Lockdown on Indian Economy

Indrajit Banerjee[1], Atul Kumar[1], Rupam Bhattacharyya[2, *]

[1]Department of Economic Studies and Policies, Central University of South Bihar

[2]Department of Biostatistics, University of Michigan

*Corresponding author. Address: 1415 Washington Heights, Ann Arbor, MI 48109, USA. Email ID: rupamb@umich.edu, Phone: +17348006834.

**Abstract**

The relationship between a pandemic and the concurrent economy is quite comparable to the relation observed among health and wealth in general. Since March 25, 2020, India had been under a nation-wide lockdown announced as a response to the spread of SARS-Cov-2 and COVID-19 and has resorted to a process of 'unlocking' the lockdown over the past couple of months. This work attempts to examine the effect of novel coronavirus 2019 (COVID-19) and its resulting disease, the COVID-19, on the foreign exchange rates and stock market performances of India using secondary data over a span of 112 days spanning between March 11 and June 30, 2020.

The study explores whether the causal relationships and directions among the growth rate of confirmed cases (GROWTHC), exchange rate (GEX) and SENSEX value (GSENSEX) are remaining the same across different pre and post-lockdown phases, attempting to capture any potential changes over time via the vector autoregressive (VAR) models. A positive correlation is found between the growth rate of confirmed cases and the growth rate of exchange rate, and a negative correlation between the growth rate of confirmed cases and the growth rate of SENSEX value.

A naïve interpretation from this could be that with the rising growth rate of the number of confirmed cases, the economy took a toll, reflected by the Indian currency being depreciated while the stock exchange index suffered from a fall. However, on applying a VAR model, it is observed that an increase in the confirmed COVID-19 cases causes no significant change in the values of the exchange rate and SENSEX index. The result varies if the analysis is split across different time periods – before lockdown, the four phases of lockdown, and the first phase of unlock.

To compare these periods, we had undertaken eight rounds of analyses. Nuanced and sensible interpretations of the numeric results indicate significant variability across time in terms of the relation between the variables of interest. This detailed knowledge about the varying patterns of dependence could potentially help the policy makers and investors of India in order to develop their strategies to cope up with the situation.

**Introduction**

A 55-year-old individual from the Hubei province in China was reported to have been the first person to have contracted what is now called COVID-19, [Bryner, 2020] the disease caused by the novel coronavirus (SARS-Cov-2). [Huang et al., 2020] In India, the first case was reported on the 30th of January 2020 in Kerala. ["COVID-19 India", 2020] Till March 24, 2020, a total of 657 COVID-19 cases were reported among which 11 were deceased and the number of recovered cases was 6. ["COVID-19 India", 2020] Since March 25, India had been under national lockdown, which extended till the end of May over four different phases. [Murukesh, 2020] The beginning of June saw India attempting to come out of the lockdown via what was christened 'unlock', a procedure that is currently at its third cycle as of August 14. [Murukesh, 2020]

One of the controlling factors of any disease scenario – quality of healthy life, which includes purified water, sanitized housing, sufficient nutritious food, good health care etc., is also an essential contributor to any economy, but the affordability of the same depends on both financial stability and access to required knowledge to enhance and maintain one's health. [Fan et al., 2018] [Farhud, 2015] [Frakt, 2018] Thus, the relationship between a pandemic and the concurrent economy has been observed to be exactly similar to the relation observed among health and wealth in general. [Bloom and Canning, 2020] [Mckee and Stuckler, 2020] [Ojong, 2020] The pandemic can hamper economic developments and trigger catastrophic events, thereby reducing the wealth of health and reducing protection against further health threats. [Bloom and Canning, 2020]

A past work on the economic impact of a pandemic observed that education, health and social service, and insurance sectors suffered from a large amount of loss to GDP in UK, France, Belgium and the Netherlands due to the effects of influenza outbreaks. [Keogh-Brown et al., 2010] According to conjectures and calculations provided by the rating agency S&P, COVID-19 can slow down the growth rate of baseline GDP for the world by 0.3 percentage point (ppt), for China by 0.7 ppt; for Asia-Pacific by 0.5 ppt; and for the USA and Europe by 0.1 to 0.2 ppt. [Bloom and Canning, 2020] Another study using data from Kuala Lumpur Stock Exchange (KLSE) and exchange rate of the currency of Malaysia (MYR) found a significant effect on equity market and on exchange rate and mentioned that the outbreak of COVID-19 is creating an insecure feeling to investors in equity market. [Bakar & Rosbi, 2020] A study on the effect of COVID-19 on financial volatility index (VIX) concluded that the spread of coronavirus is increasing the financial volatility. [Albulescu, 2020] Another study to see the short-run effect of COVID-19 on the 21 leading stock markets in the world observed a sharp decline in the stock markets since the virus outbreak. [Liu et al., 2020]

In any economy, both the supply and the demand are affected by the pandemic which may result in a persistent and large economic catastrophe all over the world. [Eichenbaum et al., 2020] However, according to McKenzie, there is no substantial relationship between Foreign Exchange volatility and international trade. [McKenzie, 1999] Vaguely, this means that at a standard scenario, the fluctuations in the foreign exchange rate do not affect the international trade. Bahmani-Oskooee and Saha investigated the potential link between exchange rate and stock prices using timeseries data of 24 countries and found that there may be a short-run relationship between these two variables, but the relationship doesn't hold in a longer term. [Bahmani-Oskooee & Saha, 2018] The short-term shocks due to a pandemic can be improved in an economy by identifying and funding productive projects with better investment opportunities which is impelled by a well-governed stock market. [Nazir et al., 2010] It has also been observed that changes in monetary policy has surprisingly variable impacts across different sectors. [Prabu et al., 2020]

In context of India, a disease outbreak and its impacts on the national economy is not a completely new experience. In 1918, "The Bombay Influenza" was spread from seven police sepoys in Bombay to Uttar Pradesh and Punjab through railway services by two weeks, which appeared to cause the highest number of deaths in one nation, the estimated number being 10-20 million. [Acharjee, 2020] During the Spanish flu, India experienced the lowest GDP growth rate (-10.5%) forever and an all-time high level of inflation, with a resulting supply side shock. The situation was worse compared to the first world war and the Bengal famine, as corroborated by future investigators. [Sreevatsan, 2020] Unsurprisingly enough, COVID-19 too presents a challenge before the Indian economy which can turn both ways. IMF predicted that the growth rate of GDP in India can be more than 7% in April 2021 if the nation can control the coronavirus outbreak, but the growth rate could come down to as low as 1.9% in March 2021, depending on the control of the disease and stimulating monetary and fiscal policies immediately ending up the lockdowns. [Choudhury, 2020]

The stock market acts as an important ingredient of an open-market economy. In this study, we consider SENSEX value to analyse the market performance of India's stock market because since 1989 when it had been set up, it has become the basic representative of the Bombay Stock Exchange (BSE) and has been considered as the 'barometer' of Indian economy. Instability of market condition results in volatility of the stock market, and the price of BSE Sensex and stock of major companies on BSE prices have previously been reported to be positively related. [Challa et al., 2018] Internationally, behaviour of experienced and smart investors has been indicated to lead to a high return. [Kaufman, 1995] Recent studies have found the market as a respondent of the potential economic outcome under the pandemic, COVID-19. [Ramelli & Wagner, 2020] Large price movements took place over the past few months because the investors started to exhibit concerns regarding the COVID-19 shock via financial ways. [Abdelnour, 2020]

This study explores the impact of COVID-19 on the stock market of India and attempts to explore their relationship using Vector Autoregressive (VAR) models. [Pfaff et al., 2008] The VAR model is treated as a standard tool in determining inter-dependencies and dynamic relationships among the variables in time-series econometrics. These models are able to explain the endogenous variables solely based on their historical values, apart from the deterministic regressors. We also aim to find out whether the fluctuations in international trade of India, caused by the pandemic, is reflected in the foreign exchange rate of the country. With new incoming data, the dependence between the economic indicators and the spread of the pandemic can be tracked more accurately, which in turn can help guide policymaking to absorb the potential diminishing effects of COVID-19.

**Methods**

Secondary data were collected for the number of infected cases from COVID-19, the stock market value and exchange rate from different sources. We have divided our work into eight Rounds considering the following eight time periods based on the decisions of government and WHO.

      a. March 25 to April 14, 2020: 21 days (Lockdown 1.0),
      b. April 15 to May 03, 2020: 19 days (Lockdown 2.0),
      c. May 04 to May 17, 2020: 14 days (Lockdown 3.0),
      d. May 18 to May 31, 2020: 14 days (Lockdown 4.0),
      e. June 01 to June 30, 2020: 30 days (Unlock 1.0),

f.  March 25 to June 30, 2020: 98 days (Lockdown 1.0 to Unlock 1.0),
g.  March 11 to April 14, 2020: 35 days (Pre-lockdown to Lockdown 1.0),
h.  March 11 to June 30: 112 days (Pre-lockdown to Unlock 1.0).

On 11$^{th}$ March, WHO upgraded the status of COVID-19 to Pandemic from Epidemic and on that day the first death in India due to coronavirus was reported. Rounds a-e are considered to identify the relations between COVID-19 on stock market and exchange rate of India exclusively for the individual phases of lockdown and unlock. Round f serves the purpose of a holistic view over the lockdown and unlock periods combined to check whether the observed patterns in the individual rounds hold for long. Round g offers the utility of taking a closer look at the dependence between the pandemic and the economic indicators at the initial, more moderate stage of the pandemic. Finally, round h is considered for 112 days to summarize the changing patterns over the whole course of the pandemic in India till the end of June.

In our analysis we use closing value of the stock market, which is significant and is a standard quantification of the market for several reasons. Investors, traders, financial institutions, regulators and other stakeholders use it as a reference point for determining performance over a specific time such as one year, a week and over a shorter time frame such as one minute or less. In fact, investors and other stakeholders base their decisions on closing stock prices. [Ellul et al., 2005] We denote the growth rates for confirmed cases, SENSEX and exchange rate as GROWTHC, GSENSEX and GEX throughout this paper. For each combination of the response and explanatory variables, we first use multiple linear regression models and then vector autoregressive (VAR) models to identify the patterns of interdependence and correlation. As diagnostic measures, we check the stationarity of the variables considered using unit root test and check the robustness of the VAR model fits using serial correlation test and lag order selection criteria. The results from the robustness checks are summarized in Supplementary Materials Section S5.

**Results**

***Round a: Lockdown 1.0 (March 25 – April 14)***

During this round, the nation was experiencing its first phase of complete lockdown due to COVID-19. The movement of the three growth rates over time for this period is summarized in Figures 1a-1c (red lines). The three variables were checked to be stationary by using unit root test (Supplementary Materials Subsection S1.1). Regression result of GEX as dependent variable on GROWTHC and GSENSEX as covariates indicates the overall model to be significant at 5% level, with the GROWTHC coefficient having a p-value of 0.08 (Supplementary Materials Subsubsection S2.1.3 and Figure 1c). One-point increase in GSENSEX causes GEX to fall by 0.058 point and the same in GROWTHC causes GEX to rise by 0.023 point (Supplementary Materials Subsubsection S2.1.3). In this period, the correlation matrix exhibits a positive correlation (0.45) between GROWTHC and GEX, but a negative correlation (-0.23) between GROWTHC and GSENSEX (Supplementary Materials Subsection S3.1). This meant that with the rise in the growth rate of the number of confirmed cases, the Indian currency continued depreciating, whereas the stock exchange index started to fall. Also, there is a notable negative correlation (-0.40) between GSENSEX and GEX. On applying VAR, we do find a significant dependence between GSENSEX and the one-period lagged difference in GROWTHC, evident by the computed t-statistic (2.16) (Supplementary Materials Subsection 4.1). Looking at the IRF graphs (Figures 1d-e), it can be seen that during this round, due to the COVID-19 shock, GSENSEX and GEX would have taken about 5.5 periods (days) to return to their long-run growth trend.

***Round b: Lockdown 2.0 (April 15 – May 03)***

During this round, the nation was experiencing the second extended phase of complete lockdown due to COVID-19. The movement of the three growth rates over time for this period is summarized in Figures 2a-2c (red lines). The three variables were checked to be stationary by using unit root test (Supplementary Materials Subsection S1.2). Regression result of GEX as dependent variable on GROWTHC and GSENSEX as covariates indicates the overall model to be significant at 5% level, with the GSENSEX coefficient having a p-value less than 0.01 (Supplementary Materials Subsubsection S2.2.3 and Figure 2c). One-point increase in GSENSEX causes GEX to fall by 0.213 point and the same in GROWTHC causes GEX to rise by 0.003 point (Supplementary Materials Subsubsection S2.2.3). In this period, the correlation matrix exhibits a weaker positive correlation (0.24) between GROWTHC and GEX compared to the same from the previous round, but a stronger negative correlation (-0.29) than before between GROWTHC and GSENSEX (Supplementary Materials Subsection S3.2). This meant that with the rise in the growth rate of the number of confirmed cases, the Indian currency still continued depreciating, whereas the stock exchange index started to fall. Also, there is a notably strong negative correlation (-0.79) between GSENSEX and GEX. On applying VAR, we find no significant dependence between the growth rates (Supplementary Materials Subsection 4.2). Looking at the IRF graphs (Figures 2d-e), it can be seen that during this round, due to the COVID-19 shock, GSENSEX and GEX would have taken about 6 and 8 periods (days) respectively to return to their long-run growth trend, indicating a larger shock.

### *Round c: Lockdown 3.0 (May 04 – May 17)*

During this round, the nation was experiencing its third extended phase of complete lockdown due to COVID-19. The movement of the three growth rates over time for this period is summarized in Figures 3a-3c (red lines). The three variables were checked to be stationary by using unit root test (Supplementary Materials Subsection S1.3). All the multiple linear regression models turn out to be non-significant at an overall 5% level. Regression result of GSENSEX as dependent variable on GROWTHC and GEX as covariates shows that ne-point increase in GEX causes GSENSEX to fall by 0.284 point and the same in GROWTHC causes GSENSEX to fall by 0.858 point (Supplementary Materials Subsubsection S2.3.2). In this period, the correlation matrix exhibits a positive correlation (0.46) between GROWTHC and GEX, but a negative correlation (-0.59) between GROWTHC and GSENSEX (Supplementary Materials Subsection S3.3). This meant that with the rise in the growth rate of the number of confirmed cases, the Indian currency continued depreciating, whereas the stock exchange index started to fall. Also, there is a notable negative correlation (-0.33) between GSENSEX and GEX. Looking at the IRF graphs from the VAR models (Figures 3d-e), it can be seen that during this round, due to the COVID-19 shock, GSENSEX and GEX would have taken about 10 periods (days) to return to their long-run growth trend.

### *Round d: Lockdown 4.0 (May 18 – May 31)*

During this round, the nation was experiencing the fourth extended phase of complete lockdown due to COVID-19. The movement of the three growth rates over time for this period is summarized in Figures 4a-4c (red lines). The three variables were checked to be stationary by using unit root test (Supplementary Materials Subsection S1.4). All the multiple linear regression models turn out to be non-significant at an overall 5% level. Regression result of GEX as dependent variable on GROWTHC and GSENSEX as covariates indicates the overall model to be significant at 10% level, with the GROWTHC coefficient having a p-value of 0.06 (Supplementary Materials Subsubsection S2.4.3 and Figure 4c). One-point increase in GSENSEX causes GEX to rise by 0.062 point and the same in GROWTHC causes GEX to rise by 0.029 point (Supplementary Materials Subsubsection S2.4.3). In this period, the correlation matrix exhibits a positive correlation (0.50) between GROWTHC and GEX, and a very small positive correlation (<0.01) between GROWTHC and GSENSEX (Supplementary Materials Subsection S3.4). This meant that with the rise in the growth rate of the number of confirmed cases, the Indian currency continued depreciating, whereas the stock exchange index experienced a very small upward effect. Also, the correlation between GSENSEX and GEX surprisingly becomes positive (0.32). Looking at the IRF graphs from the VAR model (Figures 4d-e), it can be seen that during this round, due to the COVID-19 shock, GSENSEX and GEX would have taken about 5 and 6 periods (days) respectively to return to their long-run growth trend.

### *Round e: Unlock 1.0 (June 01 – June 30)*

During this round, the nation was experiencing its first phase of unlock procedure. The movement of GSENSEX over time for this period is summarized in Figure 5a (red line). The three variables were checked for stationarity, and only GSENSEX and GEX turned out to be stationary by using unit root test (Supplementary Materials Subsection S1.5). Regression result of GSENSEX as dependent variable on GEX indicates the overall model to be non-significant at 5% level, with the GEX coefficient having a p-value greater than 0.10 (Supplementary Materials Subsubsection S2.5.1 and Figure 5a). One-point increase in GEX causes GSENSEX to fall by 1.39 points (Supplementary Materials Subsubsection S2.5.1). In this period, the correlation matrix exhibits a small positive correlation (0.12) between GROWTHC and GEX, and a larger positive correlation (0.20) between GROWTHC and GSENSEX (Supplementary Materials Subsection S3.5). This meant that with the rise in the growth rate of the number of confirmed cases, the Indian currency now started improving, and the stock exchange index started to perform better in first-order growth. Also, there is a notable negative correlation (-0.30) between GSENSEX and GEX. On applying VAR, we do find a significant dependence between GSENSEX and the one-period lagged difference in GROWTHC, evident by the computed t-statistic (2.19) (Supplementary Materials Subsection 4.5). Looking at the IRF graphs (Figures 5b-c), it can be seen that during this round, due to the COVID-19 shock, GSENSEX and GEX would have taken about 8 and 4 periods (days) respectively to return to their long-run growth trend.

### *Round f: Lockdown 1.0 to Unlock 1.0 (March 25 – June 30)*

This round was considered to observe the overall timeline of the lockdown till the end of the first unlock stage. The movement of the three growth rates over time for this period is summarized in Figures 6a-6c (red lines). The three variables were checked to be stationary by using unit root test (Supplementary Materials Subsection S1.6). Both the models with GEX and GSENSEX as responses turn out to be significant at 5% level, with both of them having extremely low (less than 0.001) p-values. Regression result of GEX as dependent variable on GROWTHC and GSENSEX as covariates indicates that one-point increase in GSENSEX causes GEX to fall by 0.078 point and the same in GROWTHC causes GEX to rise by 0.012 point (Supplementary Materials Subsubsection S2.6.3). In this period, the correlation matrix exhibits a positive correlation (0.22) between GROWTHC and GEX, but a very small negative correlation (-0.06) between GROWTHC and GSENSEX (Supplementary Materials Subsection S3.6). This meant that with the rise in the growth rate of the number of confirmed cases, the Indian currency continued improving, whereas the stock exchange index started to experience a slightly negative growth. Also, there is a notable negative correlation (-0.39) between GSENSEX and GEX. On applying VAR, we do find a significant dependence between GSENSEX till a four-period lagged difference in GROWTHC, evident by the computed t-statistics (Supplementary Materials Subsection 4.1).

### *Round g: Pre-lockdown to Lockdown 1.0 (March 11 – April 14)*

This round allows us to take a closer look at the relations between the growth rates of interest at the initial stage of the pandemic, till the end of the first lockdown. In this period, the correlation matrix exhibits a positive correlation (0.40) between GROWTHC and GEX, but a negative correlation (-0.13) between GROWTHC and GSENSEX (Supplementary Materials Subsection S3.7). This meant that with the rise in the growth rate of the number of confirmed cases, the Indian currency continued depreciating, whereas the stock exchange index started to fall. Also, there is a notable negative correlation (-0.52) between GSENSEX and GEX. Looking at the IRF graphs from the VAR model (Figures 7a-b), it can be seen that during this round, due to the COVID-19 shock, GSENSEX and GEX would have taken about 3 periods (days) to return to their long-run growth trend.

### *Round h: Pre-lockdown to Unlock 1.0 (March 11 – June 30)*

This round was considered to observe the overall timeline of the pandemic till the end of the first unlock stage. The three variables were checked to be stationary by using unit root test (Supplementary Materials Subsection S1.7). In this period, the correlation matrix exhibits a positive correlation (0.28) between GROWTHC and GEX, but a negative correlation (-0.15) between GROWTHC and GSENSEX (Supplementary Materials Subsection S3.1). This meant that with the rise in the growth rate of the number of confirmed cases, the Indian currency continued depreciating, whereas the stock exchange index started to fall. Also, there is a notable negative correlation (-0.47) between GSENSEX and GEX. On applying VAR, we do find a significant dependence between GSENSEX and the two and four-period lagged differences in GROWTHC, evident by the computed t-statistics (Supplementary Materials Subsection 4.8).

**Discussions**

The main objective of this study is to explore the relationship among the number of infected cases, exchange rate and stock market in India and to analyze the correlation among the growth rate series of these three variables. Since October 07, 2019 to March 10, 2020, average exchange rate of USD to INR was at 71.44 but after march 11, when the COVID-19 has been declared to be Pandemic by WHO, the average exchange rate over the period considered became above 75. Overall, during the course of the pandemic till the first unlock period in India, we observed a generally positive correlation between the growth rate of infected cases and the growth rate of exchange rate. This correlation was found to be negative between the growth rate of infected cases and the growth rate of Sensex.

The VAR results implied that the increase in the confirmed COVID-19 cases caused suggestive but not statistically significant changes in the values of the exchange rate and Sensex, and that due to the COVID-19 shock, the growth rate of Sensex and the growth rate of exchange rates would have taken a substantial number of days, in general, to return to their long-run growth trends. The decision of lockdown appears to be effective in controlling the spread of COVID-19 reflected by the average daily growth rate of confirmed cases and death cases. The daily average growth rate of confirmed cases was 17.804% before the lockdown which changes first to 15.684% in first phase of lockdown and then to 7.55% when the lockdown is extended. Similarly, the growth rate of deaths was 23.55% before the lockdown which first falls to 19.39% in first phase of lockdown and again to 7.27% when it is extended after April 15. These are indications that the lockdown improved the situation to some extent.

During the first phase of lockdown, the growth rate of exchange rate stays lower than the daily average growth rate of exchange rate observed before lockdown. With the extension of lockdown, the daily growth rate of exchange rate becomes negative (INR being appreciated) after April 15. During Lockdown 1.0, there is a positive correlation between GROWTHC and GEX, but a negative correlation between GROWTHC and GSENSEX. This meant that with the rise in the growth rate of the number of confirmed cases, the Indian currency continued depreciating, whereas the stock exchange index started to fall. Also, there is a strong negative correlation between the exchange rate and the Sensex index, implying that with the Indian currency depreciating the stock exchange also dropped. With extension of the lockdown, clear variabilities were observed in the relationships across the variables of interest. We found a high degree of positive correlation between GROWTHC and GEX initially, which stayed positive for most of the time over the entire period considered. The negative correlation between GROWTHC and GSENSEX was almost washed out at a point and then started showing up again towards the end.

Our investigations into the relations and interdependence between economic and health variables in context of the pandemic and the lockdown identify strong evidences of variability in these relationships over time. It is quite evident that these changes are extremely dynamic, and nuanced observation at the rapidly-changing scenario is required for effective policy decisions. Our framework is quite flexible and quick updating of the results is possible based on new incoming data, which is why it can potentially be useful in identifying the current needs from a decision-making perspective in the battle against COVID-19.

# Figures

**Figure 1: Summary of results for round a: Lockdown 1.0 (March 25 – April 14).** In panels a-c, M3 and M4 respectively indicate the months of March and April. In panels d-e, 1 in the x-axis indicates the beginning of the period considered.

**Panel a:** Summaries corresponding to regression of GROWTHC on GSENSEX and GEX where the blue line, the red line and the green line represent residual, actual and fitted values respectively.

**Panel b:** Summaries corresponding to regression of GSENSEX on GROWTHC and GEX where the blue line, the red line and the green line represent residual, actual and fitted values respectively.

**Panel c:** Summaries corresponding to regression of GEX on GSENSEX and GROWTHC where the blue line, the red line and the green line represent residual, actual and fitted values respectively.

**Panel d:** Impulse Response Function graph of GSENSEX to GROWTHC obtained from the fitted vector autoregressive model.

**Panel e:** Impulse Response Function graph of GEX to GROWTHC obtained from the fitted vector autoregressive model.

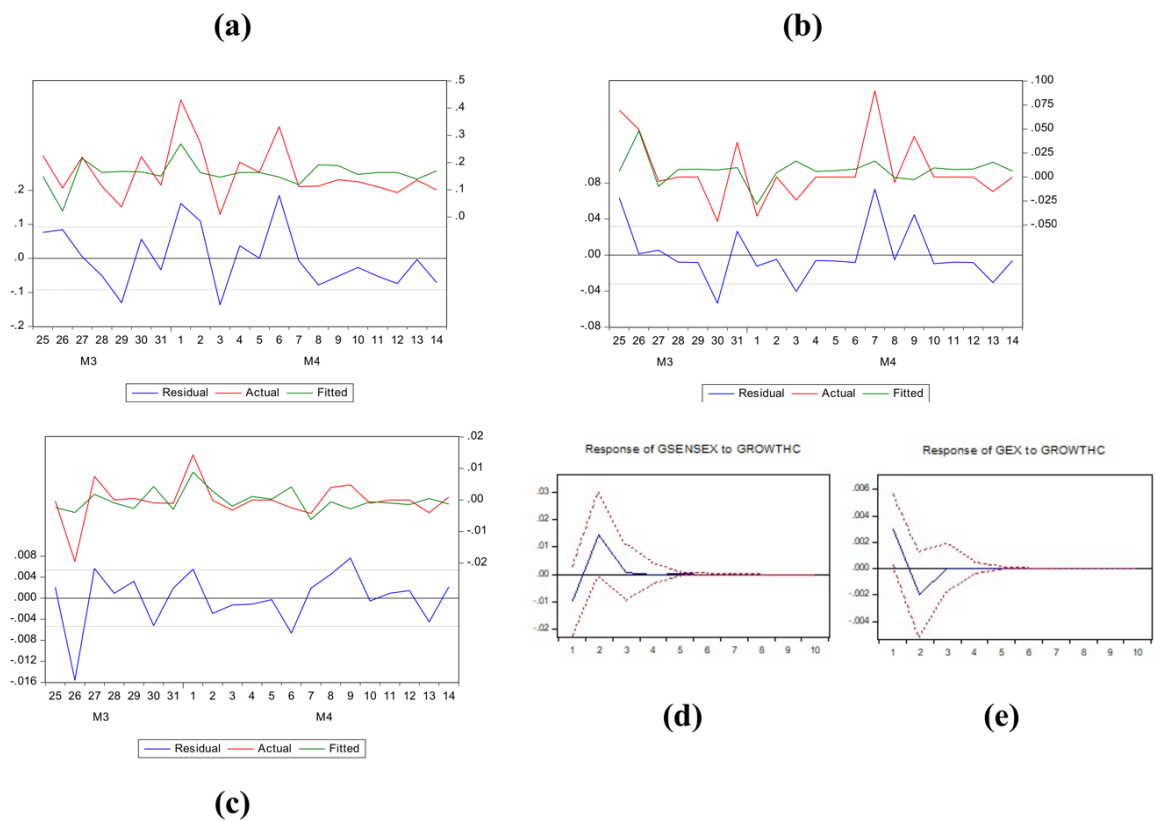

**Figure 2: Summary of results for round b: Lockdown 2.0 (April 15 – May 03).** In panels a-c, 2020m4 and 2020m5 respectively indicate the months of April and May. In panels d-e, 1 in the x-axis indicates the beginning of the period considered.

**Panel a:** Summaries corresponding to regression of GROWTHC on GSENSEX and GEX where the blue line, the red line and the green line represent residual, actual and fitted values respectively.

**Panel b:** Summaries corresponding to regression of GSENSEX on GROWTHC and GEX where the blue line, the red line and the green line represent residual, actual and fitted values respectively.

**Panel c:** Summaries corresponding to regression of GEX on GSENSEX and GROWTHC where the blue line, the red line and the green line represent residual, actual and fitted values respectively.

**Panel d:** Impulse Response Function graph of GSENSEX to GROWTHC obtained from the fitted vector autoregressive model.

**Panel e:** Impulse Response Function graph of GEX to GROWTHC obtained from the fitted vector autoregressive model.

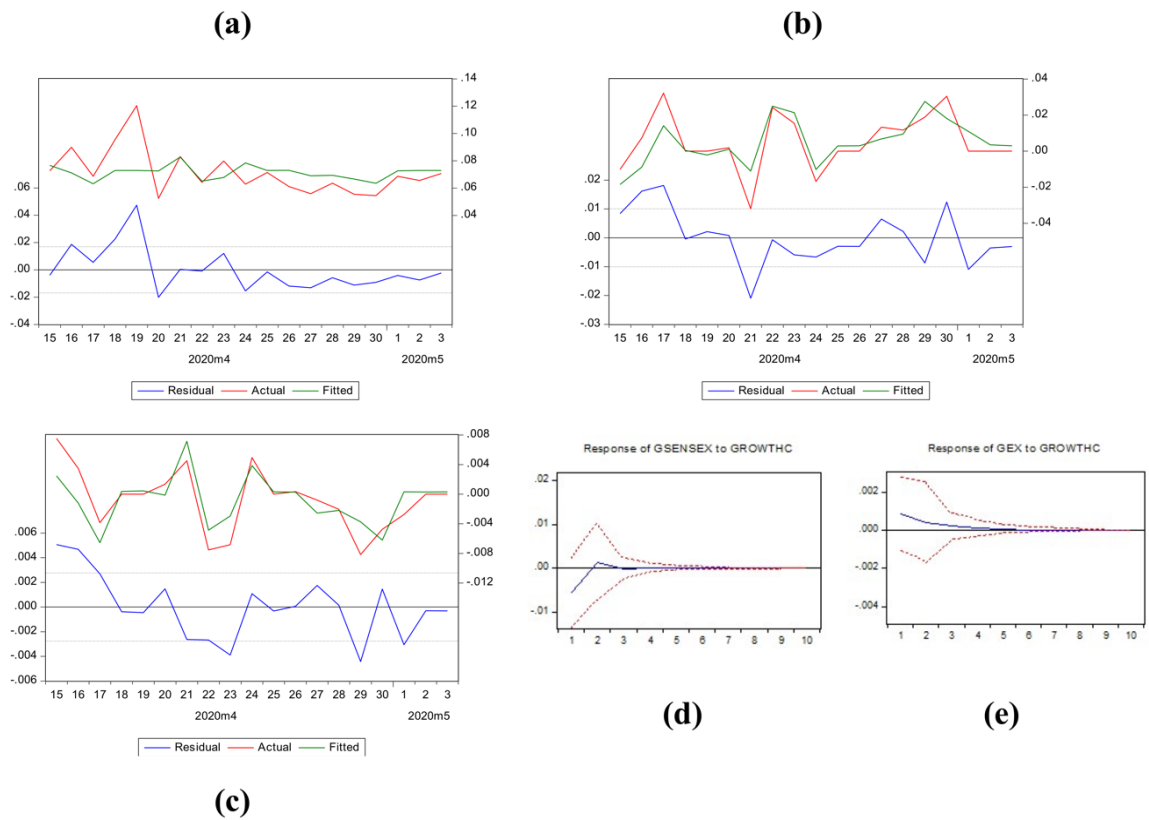

**Figure 3: Summary of results for round c: Lockdown 3.0 (May 04 – May 17).** In panels a-c, 2020m5 indicates the month of May. In panels d-e, 1 in the x-axis indicates the beginning of the period considered.

**Panel a:** Summaries corresponding to regression of GROWTHC on GSENSEX and GEX where the blue line, the red line and the green line represent residual, actual and fitted values respectively.

**Panel b:** Summaries corresponding to regression of GSENSEX on GROWTHC and GEX where the blue line, the red line and the green line represent residual, actual and fitted values respectively.

**Panel c:** Summaries corresponding to regression of GEX on GSENSEX and GROWTHC where the blue line, the red line and the green line represent residual, actual and fitted values respectively.

**Panel d:** Impulse Response Function graph of GSENSEX to GROWTHC obtained from the fitted vector autoregressive model.

**Panel e:** Impulse Response Function graph of GEX to GROWTHC obtained from the fitted vector autoregressive model.

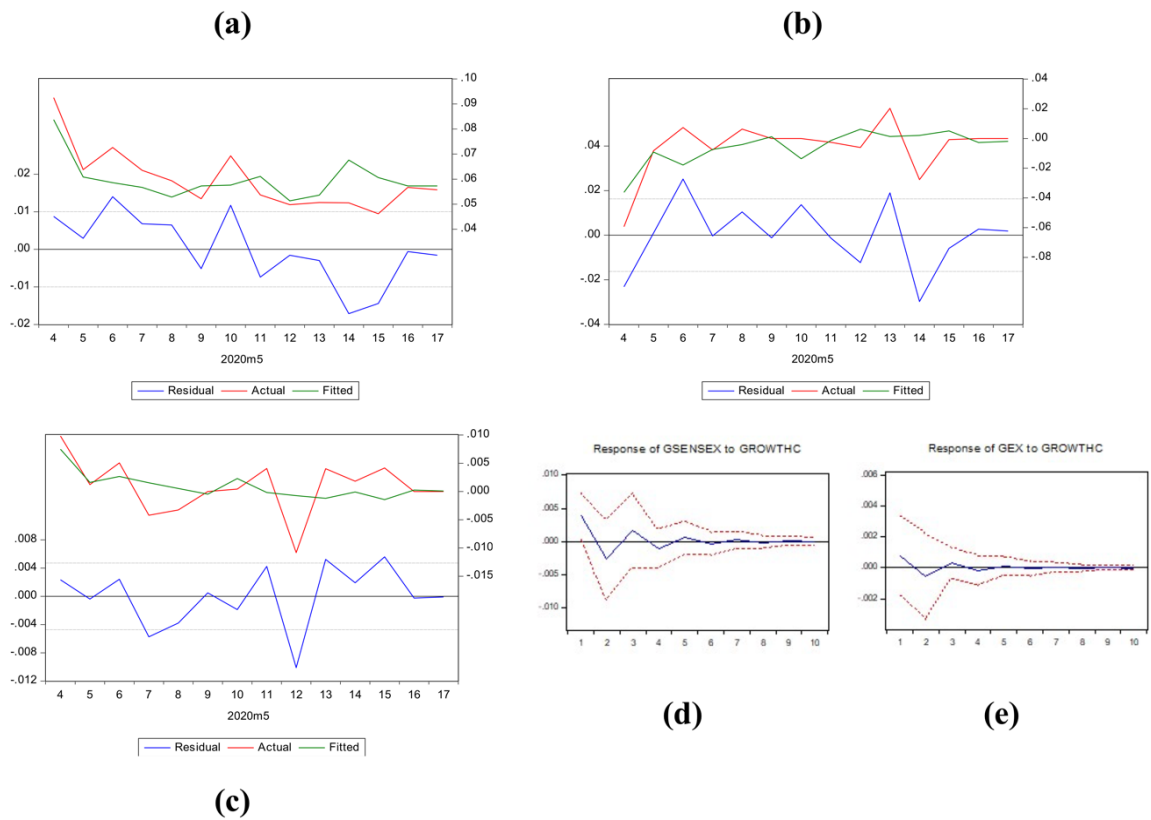

**Figure 4: Summary of results for round d: Lockdown 4.0 (May 18 – May 31).** In panels a-c, 2020m5 indicates the month of May. In panels d-e, 1 in the x-axis indicates the beginning of the period considered.

**Panel a:** Summaries corresponding to regression of GROWTHC on GSENSEX and GEX where the blue line, the red line and the green line represent residual, actual and fitted values respectively.

**Panel b:** Summaries corresponding to regression of GSENSEX on GROWTHC and GEX where the blue line, the red line and the green line represent residual, actual and fitted values respectively.

**Panel c:** Summaries corresponding to regression of GEX on GSENSEX and GROWTHC where the blue line, the red line and the green line represent residual, actual and fitted values respectively.

**Panel d:** Impulse Response Function graph of GSENSEX to GROWTHC obtained from the fitted vector autoregressive model.

**Panel e:** Impulse Response Function graph of GEX to GROWTHC obtained from the fitted vector autoregressive model.

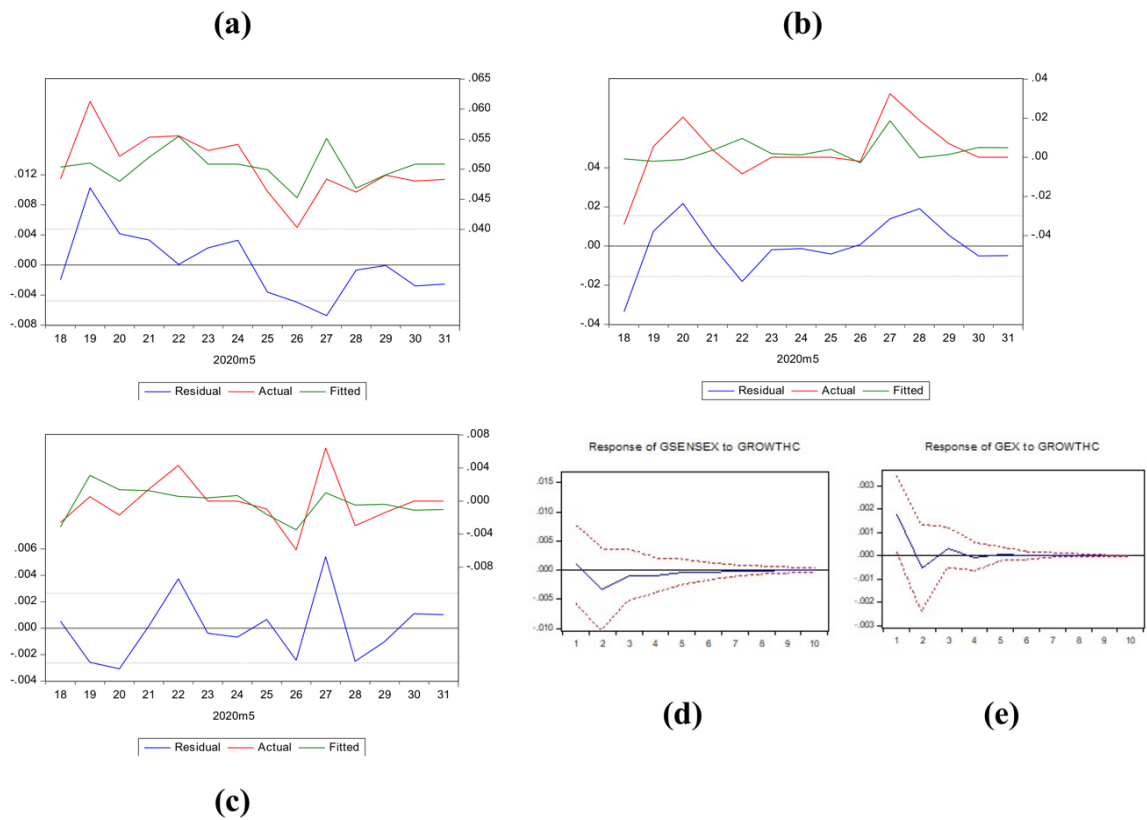

**Figure 5: Summary of results for round a: Unlock 1.0 (June 01 – June 30).** In panel a, M6 indicates the month of June. In panels b-c, 1 in the x-axis indicates the beginning of the period considered.

**Panel a:** Summaries corresponding to regression of GSENSEX on GROWTHC and GEX where the blue line, the red line and the green line represent residual, actual and fitted values respectively.

**Panel b:** Impulse Response Function graph of GSENSEX to D(GROWTHC) obtained from the fitted vector autoregressive model.

**Panel c:** Impulse Response Function graph of GEX to D(GROWTHC) obtained from the fitted vector autoregressive model.

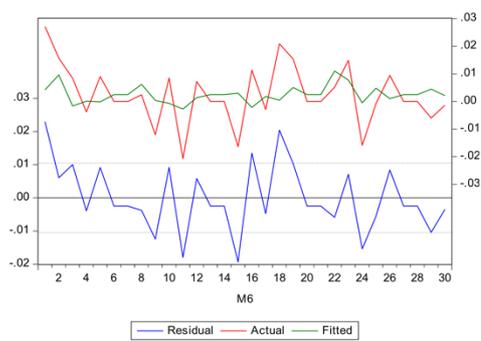
(a)

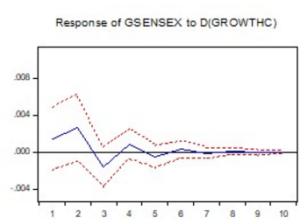
(b)

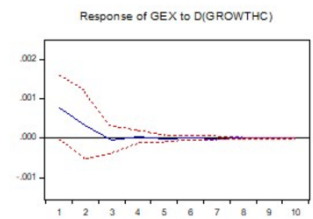
(c)

**Figure 6: Summary of results for round f: Lockdown 1.0 to Unlock 1.0 (March 25 – June 30).** In panels a-c, M3, M4, M5, and M6 respectively indicate the months of March, April, May, and June. In panels d-e, 1 in the x-axis indicates the beginning of the period considered.

**Panel a:** Summaries corresponding to regression of GROWTHC on GSENSEX and GEX where the blue line, the red line and the green line represent residual, actual and fitted values respectively.

**Panel b:** Summaries corresponding to regression of GSENSEX on GROWTHC and GEX where the blue line, the red line and the green line represent residual, actual and fitted values respectively.

**Panel c:** Summaries corresponding to regression of GEX on GSENSEX and GROWTHC where the blue line, the red line and the green line represent residual, actual and fitted values respectively.

**Panel d:** Impulse Response Function graph of GSENSEX to GROWTHC obtained from the fitted vector autoregressive model.

**Panel e:** Impulse Response Function graph of GEX to GROWTHC obtained from the fitted vector autoregressive model.

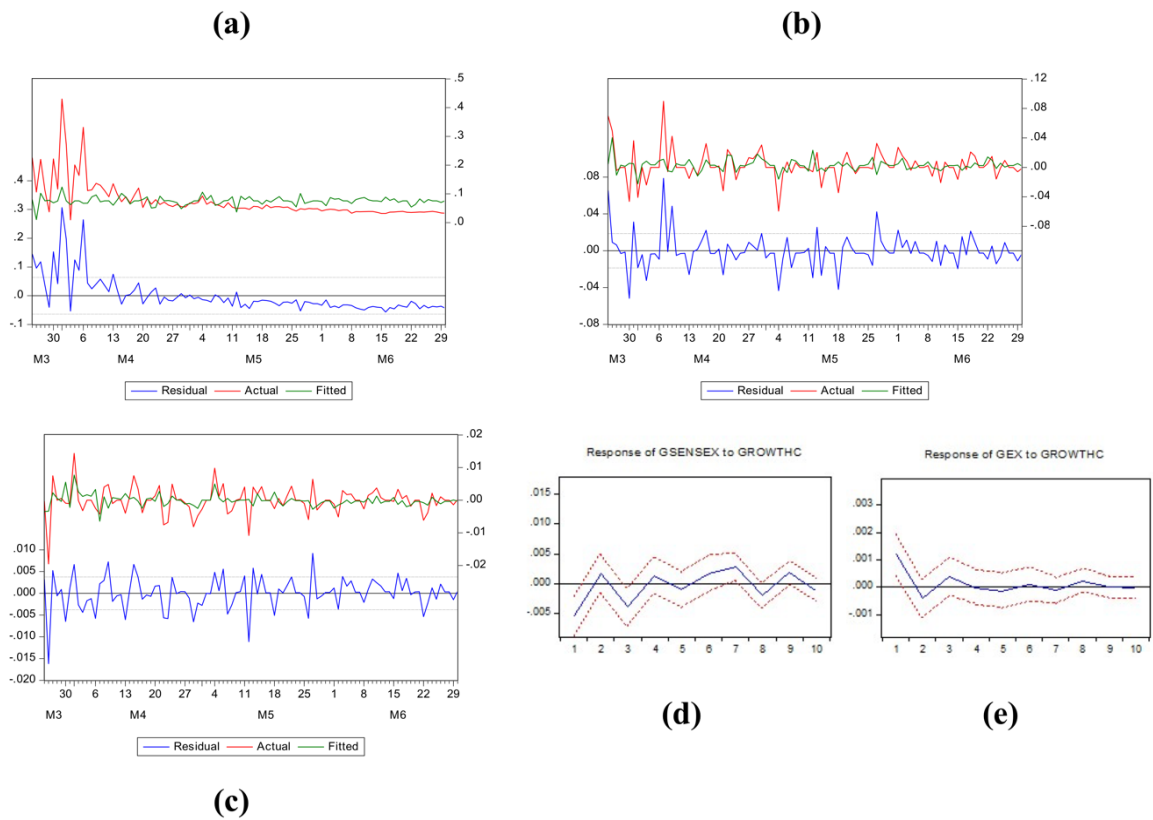

**Figure 7: Summary of results for round g: Pre-lockdown to Lockdown 1.0 (March 11 – April 14).** In panels a-b, 1 in the x-axis indicates the beginning of the period considered.

**Panel a:** Impulse Response Function graph of GSENSEX to GROWTHC obtained from the fitted vector autoregressive model.

**Panel b:** Impulse Response Function graph of GEX to GROWTHC obtained from the fitted vector autoregressive model.

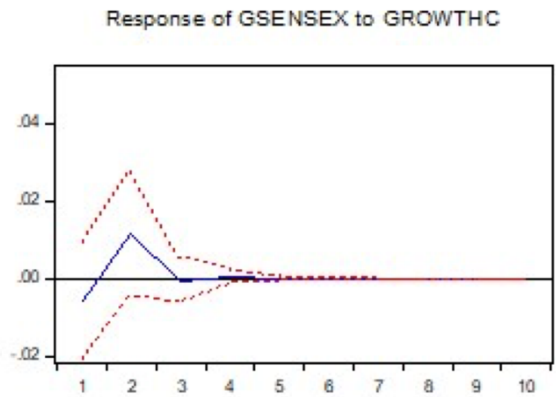
(a)

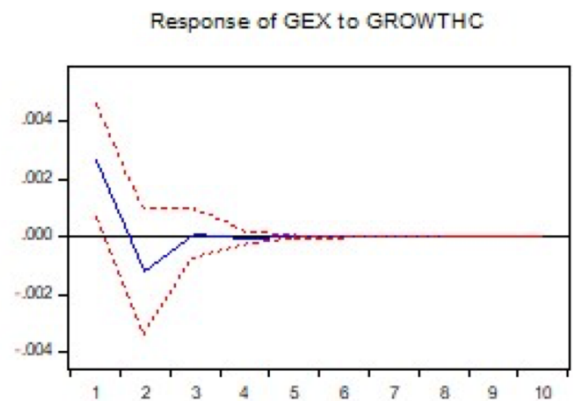
(b)

**Figure 8: Summary of results for round h: Pre-lockdown to Unlock 1.0 (March 11 – June 30).** In panels a-b, 1 in the x-axis indicates the beginning of the period considered.

**Panel a:** Impulse Response Function graph of GSENSEX to GROWTHC obtained from the fitted vector autoregressive model.

**Panel b:** Impulse Response Function graph of GEX to GROWTHC obtained from the fitted vector autoregressive model.

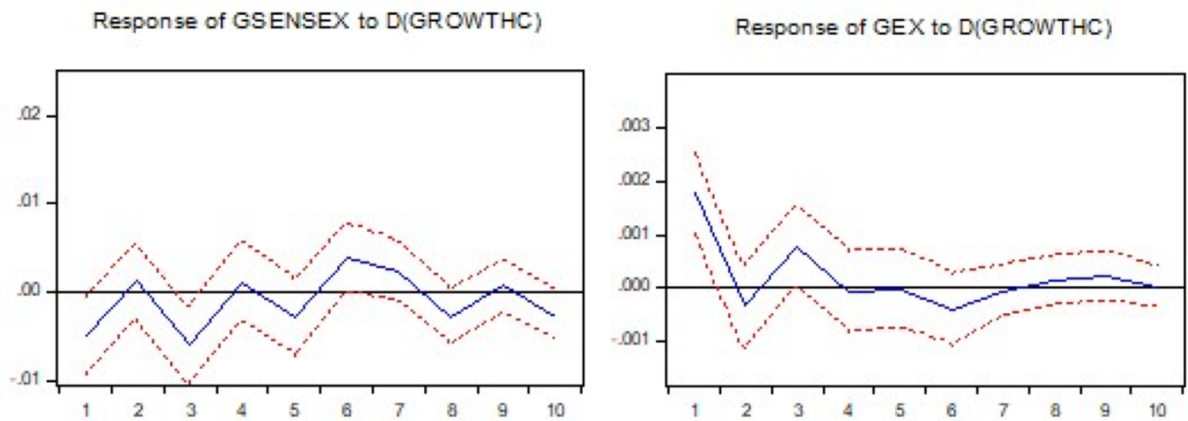

(a)          (b)

**Supplementary Document for**

**Examining the Effect of COVID-19 on Foreign Exchange Rate and Stock Market – An Applied Insight into the Variable Effects of Lockdown on Indian Economy**

Indrajit Banerjee[1], Atul Kumar[1], Rupam Bhattacharyya[2, *]

[1]Department of Economic Studies and Policies, Central University of South Bihar

[2]Department of Biostatistics, University of Michigan

*Corresponding author. Address: 1415 Washington Heights, Ann Arbor, MI 48109, USA. Email ID: rupamb@umich.edu, Phone: +17348006834.

**Supplementary Materials**

*S1. Results of unit-root tests across different rounds*

*S1.1. Round a: Lockdown 1.0 (March 25 – April 14)*

*S1.1.1. Unit-root test for GROWTHC*

Null Hypothesis: GROWTHC has a unit root
Exogenous: Constant
Lag Length: 0 (Automatic - based on SIC, maxlag=4)

|  |  | t-Statistic | Prob.* |
|---|---|---|---|
| Augmented Dickey-Fuller test statistic |  | -4.566172 | 0.0020 |
| Test critical values: | 1% level | -3.808546 |  |
|  | 5% level | -3.020686 |  |
|  | 10% level | -2.650413 |  |

*MacKinnon (1996) one-sided p-values.

Augmented Dickey-Fuller Test Equation
Dependent Variable: D(GROWTHC)
Method: Least Squares
Date: 07/08/20   Time: 12:22
Sample (adjusted): 3/26/2020 4/14/2020
Included observations: 20 after adjustments

| Variable | Coefficient | Std. Error | t-Statistic | Prob. |
|---|---|---|---|---|
| GROWTHC(-1) | -1.072300 | 0.234836 | -4.566172 | 0.0002 |
| C | 0.169459 | 0.044689 | 3.791935 | 0.0013 |
| R-squared | 0.536679 | Mean dependent var | -0.006341 | |
| Adjusted R-squared | 0.510939 | S.D. dependent var | 0.145101 | |
| S.E. of regression | 0.101473 | Akaike info criterion | -1.643401 | |
| Sum squared resid | 0.185343 | Schwarz criterion | -1.543828 | |
| Log likelihood | 18.43401 | Hannan-Quinn criter. | -1.623963 | |
| F-statistic | 20.84993 | Durbin-Watson stat | 1.945374 | |
| Prob(F-statistic) | 0.000239 | | | |

### *S1.1.2. Unit-root test for GSENSEX*

Null Hypothesis: GSENSEX has a unit root

Exogenous: None

Lag Length: 0 (Automatic - based on SIC, maxlag=4)

| | | t-Statistic | Prob.* |
|---|---|---|---|
| Augmented Dickey-Fuller test statistic | | -5.051439 | 0.0000 |
| Test critical values: | 1% level | -2.685718 | |
| | 5% level | -1.959071 | |
| | 10% level | -1.607456 | |

*MacKinnon (1996) one-sided p-values.

Augmented Dickey-Fuller Test Equation

Dependent Variable: D(GSENSEX)

Method: Least Squares

Date: 07/08/20   Time: 12:23

Sample (adjusted): 3/26/2020 4/14/2020

Included observations: 20 after adjustments

| Variable | Coefficient | Std. Error | t-Statistic | Prob. |
|---|---|---|---|---|

| | | | |
|---|---|---|---|
| GSENSEX(-1) | -1.028991 | 0.203703 | -5.051439 | 0.0001 |

| | | | |
|---|---|---|---|
| R-squared | 0.570748 | Mean dependent var | -0.003490 |
| Adjusted R-squared | 0.570748 | S.D. dependent var | 0.047269 |
| S.E. of regression | 0.030970 | Akaike info criterion | -4.062912 |
| Sum squared resid | 0.018223 | Schwarz criterion | -4.013126 |
| Log likelihood | 41.62912 | Hannan-Quinn criter. | -4.053193 |
| Durbin-Watson stat | 2.228737 | | |

*S1.1.3. Unit-root test for GEX*

Null Hypothesis: GEX has a unit root

Exogenous: None

Lag Length: 0 (Automatic - based on SIC, maxlag=4)

| | | t-Statistic | Prob.* |
|---|---|---|---|
| Augmented Dickey-Fuller test statistic | | -5.370109 | 0.0000 |
| Test critical values: | 1% level | -2.685718 | |
| | 5% level | -1.959071 | |
| | 10% level | -1.607456 | |

*MacKinnon (1996) one-sided p-values.

Augmented Dickey-Fuller Test Equation

Dependent Variable: D(GEX)

Method: Least Squares

Date: 07/08/20   Time: 12:24

Sample (adjusted): 3/26/2020 4/14/2020

Included observations: 20 after adjustments

| Variable | Coefficient | Std. Error | t-Statistic | Prob. |
|---|---|---|---|---|
| GEX(-1) | -1.206194 | 0.224613 | -5.370109 | 0.0000 |

| | | | |
|---|---|---|---|
| R-squared | 0.602810 | Mean dependent var | 6.10E-05 |
| Adjusted R-squared | 0.602810 | S.D. dependent var | 0.009668 |
| S.E. of regression | 0.006093 | Akaike info criterion | -7.314523 |
| Sum squared resid | 0.000705 | Schwarz criterion | -7.264737 |
| Log likelihood | 74.14523 | Hannan-Quinn criter. | -7.304805 |
| Durbin-Watson stat | 1.544222 | | |

### *S1.2. Round b: Lockdown 2.0 (April 15 – May 03)*

### *S1.2.1. Unit-root test for GROWTHC*

Null Hypothesis: GROWTHC has a unit root

Exogenous: Constant

Lag Length: 0 (Automatic - based on SIC, maxlag=3)

| | | t-Statistic | Prob.* |
|---|---|---|---|
| Augmented Dickey-Fuller test statistic | | -3.650822 | 0.0152 |
| Test critical values: | 1% level | -3.857386 | |
| | 5% level | -3.040391 | |
| | 10% level | -2.660551 | |

*MacKinnon (1996) one-sided p-values.

Warning: Probabilities and critical values calculated for 20 observations and may not be accurate for a sample size of 18

Augmented Dickey-Fuller Test Equation

Dependent Variable: D(GROWTHC)

Method: Least Squares

Date: 07/08/20   Time: 12:26

Sample (adjusted): 4/16/2020 5/03/2020

Included observations: 18 after adjustments

| Variable | Coefficient | Std. Error | t-Statistic | Prob. |
|---|---|---|---|---|
| GROWTHC(-1) | -0.908795 | 0.248929 | -3.650822 | 0.0022 |
| C | 0.064823 | 0.018265 | 3.549110 | 0.0027 |

| | | | |
|---|---|---|---|
| R-squared | 0.454456 | Mean dependent var | -0.000112 |
| Adjusted R-squared | 0.420359 | S.D. dependent var | 0.023140 |
| S.E. of regression | 0.017617 | Akaike info criterion | -5.135438 |
| Sum squared resid | 0.004966 | Schwarz criterion | -5.036508 |
| Log likelihood | 48.21894 | Hannan-Quinn criter. | -5.121797 |
| F-statistic | 13.32850 | Durbin-Watson stat | 1.991684 |
| Prob(F-statistic) | 0.002155 | | |

*S1.2.2. Unit-root test for GSENSEX*

Null Hypothesis: GSENSEX has a unit root

Exogenous: None

Lag Length: 0 (Automatic - based on SIC, maxlag=3)

| | | t-Statistic | Prob.* |
|---|---|---|---|
| Augmented Dickey-Fuller test statistic | | -3.846746 | 0.0007 |
| Test critical values: | 1% level | -2.699769 | |
| | 5% level | -1.961409 | |
| | 10% level | -1.606610 | |

*MacKinnon (1996) one-sided p-values.

Warning: Probabilities and critical values calculated for 20 observations and may not be accurate for a sample size of 18

Augmented Dickey-Fuller Test Equation

Dependent Variable: D(GSENSEX)

Method: Least Squares

Date: 07/08/20   Time: 12:27

Sample (adjusted): 4/16/2020 5/03/2020

Included observations: 18 after adjustments

| Variable | Coefficient | Std. Error | t-Statistic | Prob. |
|---|---|---|---|---|
| GSENSEX(-1) | -0.920242 | 0.239226 | -3.846746 | 0.0013 |

| | | | |
|---|---|---|---|
| R-squared | 0.465027 | Mean dependent var | 0.000562 |
| Adjusted R-squared | 0.465027 | S.D. dependent var | 0.022954 |
| S.E. of regression | 0.016789 | Akaike info criterion | -5.282262 |
| Sum squared resid | 0.004792 | Schwarz criterion | -5.232797 |
| Log likelihood | 48.54036 | Hannan-Quinn criter. | -5.275441 |
| Durbin-Watson stat | 1.931701 | | |

*S1.2.3. Unit-root test for GEX*

Null Hypothesis: GEX has a unit root

Exogenous: None

Lag Length: 0 (Automatic - based on SIC, maxlag=3)

| | | t-Statistic | Prob.* |
|---|---|---|---|
| Augmented Dickey-Fuller test statistic | | -3.652402 | 0.0011 |
| Test critical values: | 1% level | -2.699769 | |
| | 5% level | -1.961409 | |
| | 10% level | -1.606610 | |

*MacKinnon (1996) one-sided p-values.

Warning: Probabilities and critical values calculated for 20 observations and may not be accurate for a sample size of 18

Augmented Dickey-Fuller Test Equation

Dependent Variable: D(GEX)

Method: Least Squares

Date: 07/08/20   Time: 12:27

Sample (adjusted): 4/16/2020 5/03/2020

Included observations: 18 after adjustments

| Variable | Coefficient | Std. Error | t-Statistic | Prob. |
|---|---|---|---|---|
| GEX(-1) | -0.784460 | 0.214779 | -3.652402 | 0.0020 |

| | | | |
|---|---|---|---|
| R-squared | 0.435893 | Mean dependent var | -0.000419 |
| Adjusted R-squared | 0.435893 | S.D. dependent var | 0.005236 |
| S.E. of regression | 0.003932 | Akaike info criterion | -8.185261 |
| Sum squared resid | 0.000263 | Schwarz criterion | -8.135796 |
| Log likelihood | 74.66735 | Hannan-Quinn criter. | -8.178440 |
| Durbin-Watson stat | 1.957083 | | |

### *S1.3. Round c: Lockdown 3.0 (May 04 – May 17)*

#### *S1.3.1. Unit-root test for GROWTHC*

Null Hypothesis: GROWTHC has a unit root

Exogenous: Constant

Lag Length: 0 (Automatic - based on SIC, maxlag=2)

| | | t-Statistic | Prob.* |
|---|---|---|---|
| Augmented Dickey-Fuller test statistic | | -4.185030 | 0.0081 |
| Test critical values: | 1% level | -4.057910 | |
| | 5% level | -3.119910 | |
| | 10% level | -2.701103 | |

*MacKinnon (1996) one-sided p-values.

Warning: Probabilities and critical values calculated for 20 observations and may not be accurate for a sample size of 13

Augmented Dickey-Fuller Test Equation

Dependent Variable: D(GROWTHC)

Method: Least Squares

Date: 07/08/20   Time: 12:29

Sample (adjusted): 5/05/2020 5/17/2020

Included observations: 13 after adjustments

| Variable | Coefficient | Std. Error | t-Statistic | Prob. |
|---|---|---|---|---|
| GROWTHC(-1) | -0.723950 | 0.172986 | -4.185030 | 0.0015 |
| C | 0.040648 | 0.010599 | 3.835088 | 0.0028 |

| | | | |
|---|---|---|---|
| R-squared | 0.614231 | Mean dependent var | -0.002831 |
| Adjusted R-squared | 0.579161 | S.D. dependent var | 0.011666 |
| S.E. of regression | 0.007568 | Akaike info criterion | -6.789090 |
| Sum squared resid | 0.000630 | Schwarz criterion | -6.702174 |
| Log likelihood | 46.12908 | Hannan-Quinn criter. | -6.806955 |
| F-statistic | 17.51447 | Durbin-Watson stat | 2.233425 |
| Prob(F-statistic) | 0.001524 | | |

*S1.3.2. Unit-root test for GSENSEX*

Null Hypothesis: GSENSEX has a unit root

Exogenous: None

Lag Length: 0 (Automatic - based on SIC, maxlag=2)

| | t-Statistic | Prob.* |
|---|---|---|
| Augmented Dickey-Fuller test statistic | -6.894809 | 0.0000 |

| Test critical values: | 1% level | -2.754993 |
| --- | --- | --- |
| | 5% level | -1.970978 |
| | 10% level | -1.603693 |

*MacKinnon (1996) one-sided p-values.

Warning: Probabilities and critical values calculated for 20 observations and may not be accurate for a sample size of 13

Augmented Dickey-Fuller Test Equation

Dependent Variable: D(GSENSEX)

Method: Least Squares

Date: 07/08/20   Time: 12:30

Sample (adjusted): 5/05/2020 5/17/2020

Included observations: 13 after adjustments

| Variable | Coefficient | Std. Error | t-Statistic | Prob. |
| --- | --- | --- | --- | --- |
| GSENSEX(-1) | -1.064886 | 0.154448 | -6.894809 | 0.0000 |

| | | | |
| --- | --- | --- | --- |
| R-squared | 0.790394 | Mean dependent var | 0.004568 |
| Adjusted R-squared | 0.790394 | S.D. dependent var | 0.023782 |
| S.E. of regression | 0.010888 | Akaike info criterion | -6.128461 |
| Sum squared resid | 0.001423 | Schwarz criterion | -6.085004 |
| Log likelihood | 40.83500 | Hannan-Quinn criter. | -6.137394 |
| Durbin-Watson stat | 2.930780 | | |

### S1.3.3. Unit-root test for GEX

Null Hypothesis: GEX has a unit root

Exogenous: None

Lag Length: 0 (Automatic - based on SIC, maxlag=2)

| | t-Statistic | Prob.* |
| --- | --- | --- |

| | | | |
|---|---|---|---|
| Augmented Dickey-Fuller test statistic | | -5.042100 | 0.0001 |
| Test critical values: | 1% level | -2.754993 | |
| | 5% level | -1.970978 | |
| | 10% level | -1.603693 | |

*MacKinnon (1996) one-sided p-values.

Warning: Probabilities and critical values calculated for 20 observations and may not be accurate for a sample size of 13

Augmented Dickey-Fuller Test Equation

Dependent Variable: D(GEX)

Method: Least Squares

Date: 07/08/20   Time: 12:30

Sample (adjusted): 5/05/2020 5/17/2020

Included observations: 13 after adjustments

| Variable | Coefficient | Std. Error | t-Statistic | Prob. |
|---|---|---|---|---|
| GEX(-1) | -1.187481 | 0.235513 | -5.042100 | 0.0003 |

| | | | |
|---|---|---|---|
| R-squared | 0.675744 | Mean dependent var | -0.000756 |
| Adjusted R-squared | 0.675744 | S.D. dependent var | 0.007430 |
| S.E. of regression | 0.004231 | Akaike info criterion | -8.018989 |
| Sum squared resid | 0.000215 | Schwarz criterion | -7.975532 |
| Log likelihood | 53.12343 | Hannan-Quinn criter. | -8.027922 |
| Durbin-Watson stat | 2.193739 | | |

### *S1.4. Round d: Lockdown 4.0 (May 18 – May 31)*

#### *S1.4.1. Unit-root test for GROWTHC*

Null Hypothesis: GROWTHC has a unit root

Exogenous: Constant, Linear Trend

Lag Length: 0 (Automatic - based on SIC, maxlag=2)

|  |  | t-Statistic | Prob.* |
|---|---|---|---|
| Augmented Dickey-Fuller test statistic |  | -3.858449 | 0.0478 |
| Test critical values: | 1% level | -4.886426 |  |
|  | 5% level | -3.828975 |  |
|  | 10% level | -3.362984 |  |

*MacKinnon (1996) one-sided p-values.

Warning: Probabilities and critical values calculated for 20 observations and may not be accurate for a sample size of 13

Augmented Dickey-Fuller Test Equation

Dependent Variable: D(GROWTHC)

Method: Least Squares

Date: 07/08/20   Time: 12:32

Sample (adjusted): 5/19/2020 5/31/2020

Included observations: 13 after adjustments

| Variable | Coefficient | Std. Error | t-Statistic | Prob. |
|---|---|---|---|---|
| GROWTHC(-1) | -1.069484 | 0.277180 | -3.858449 | 0.0032 |
| C | 0.061259 | 0.015776 | 3.882963 | 0.0030 |
| @TREND("5/18/2020") | -0.001017 | 0.000381 | -2.667005 | 0.0236 |

| R-squared | 0.602591 | Mean dependent var | -6.36E-06 |
|---|---|---|---|
| Adjusted R-squared | 0.523110 | S.D. dependent var | 0.006069 |
| S.E. of regression | 0.004191 | Akaike info criterion | -7.912626 |
| Sum squared resid | 0.000176 | Schwarz criterion | -7.782253 |
| Log likelihood | 54.43207 | Hannan-Quinn criter. | -7.939423 |
| F-statistic | 7.581512 | Durbin-Watson stat | 1.356884 |

| | | |
|---|---|---|
| Prob(F-statistic) | 0.009913 | |

*S1.4.2. Unit-root test for GSENSEX*

Null Hypothesis: GSENSEX has a unit root

Exogenous: None

Lag Length: 0 (Automatic - based on SIC, maxlag=2)

| | | t-Statistic | Prob.* |
|---|---|---|---|
| Augmented Dickey-Fuller test statistic | | -3.601657 | 0.0017 |
| Test critical values: | 1% level | -2.754993 | |
| | 5% level | -1.970978 | |
| | 10% level | -1.603693 | |

*MacKinnon (1996) one-sided p-values.

Warning: Probabilities and critical values calculated for 20 observations and may not be accurate for a sample size of 13

Augmented Dickey-Fuller Test Equation

Dependent Variable: D(GSENSEX)

Method: Least Squares

Date: 07/08/20   Time: 12:32

Sample (adjusted): 5/19/2020 5/31/2020

Included observations: 13 after adjustments

| Variable | Coefficient | Std. Error | t-Statistic | Prob. |
|---|---|---|---|---|
| GSENSEX(-1) | -0.797588 | 0.221450 | -3.601657 | 0.0036 |

| | | | |
|---|---|---|---|
| R-squared | 0.508009 | Mean dependent var | 0.002644 |
| Adjusted R-squared | 0.508009 | S.D. dependent var | 0.017825 |
| S.E. of regression | 0.012503 | Akaike info criterion | -5.851947 |
| Sum squared resid | 0.001876 | Schwarz criterion | -5.808489 |
| Log likelihood | 39.03765 | Hannan-Quinn criter. | -5.860879 |
| Durbin-Watson stat | 1.283765 | | |

### S1.4.3. Unit-root test for GEX

Null Hypothesis: GEX has a unit root

Exogenous: None

Lag Length: 0 (Automatic - based on SIC, maxlag=2)

| | | t-Statistic | Prob.* |
|---|---|---|---|
| Augmented Dickey-Fuller test statistic | | -5.392857 | 0.0001 |
| Test critical values: | 1% level | -2.754993 | |
| | 5% level | -1.970978 | |
| | 10% level | -1.603693 | |

*MacKinnon (1996) one-sided p-values.

Warning: Probabilities and critical values calculated for 20 observations and may not be accurate for a sample size of 13

Augmented Dickey-Fuller Test Equation

Dependent Variable: D(GEX)

Method: Least Squares

Date: 07/08/20   Time: 12:33

Sample (adjusted): 5/19/2020 5/31/2020

Included observations: 13 after adjustments

| Variable | Coefficient | Std. Error | t-Statistic | Prob. |
|---|---|---|---|---|
| GEX(-1) | -1.386764 | 0.257148 | -5.392857 | 0.0002 |
| R-squared | 0.707436 | Mean dependent var | | 0.000200 |
| Adjusted R-squared | 0.707436 | S.D. dependent var | | 0.005171 |

| | | |
|---|---|---|
| S.E. of regression | 0.002797 | Akaike info criterion -8.846919 |
| Sum squared resid | 9.39E-05 | Schwarz criterion -8.803461 |
| Log likelihood | 58.50497 | Hannan-Quinn criter. -8.855852 |
| Durbin-Watson stat | 2.166973 | |

### S1.5. Round e: Unlock 1.0 (June 01 – June 30)

#### S1.5.1. Unit-root test for GROWTHC

Null Hypothesis: D(GROWTHC) has a unit root

Exogenous: Constant, Linear Trend

Lag Length: 0 (Automatic - based on SIC, maxlag=7)

| | | t-Statistic | Prob.* |
|---|---|---|---|
| Augmented Dickey-Fuller test statistic | | -6.131553 | 0.0001 |
| Test critical values: | 1% level | -4.323979 | |
| | 5% level | -3.580623 | |
| | 10% level | -3.225334 | |

*MacKinnon (1996) one-sided p-values.

Augmented Dickey-Fuller Test Equation

Dependent Variable: D(GROWTHC,2)

Method: Least Squares

Date: 07/08/20   Time: 12:39

Sample (adjusted): 6/03/2020 6/30/2020

Included observations: 28 after adjustments

| Variable | Coefficient | Std. Error | t-Statistic | Prob. |
|---|---|---|---|---|
| D(GROWTHC(-1)) | -1.170694 | 0.190929 | -6.131553 | 0.0000 |
| C | -0.000570 | 0.001281 | -0.444976 | 0.6602 |
| @TREND("6/01/2020") | 7.36E-06 | 7.34E-05 | 0.100304 | 0.9209 |

| | | | |
|---|---|---|---|
| R-squared | 0.601475 | Mean dependent var | -0.000166 |
| Adjusted R-squared | 0.569593 | S.D. dependent var | 0.004775 |
| S.E. of regression | 0.003132 | Akaike info criterion | -8.593100 |
| Sum squared resid | 0.000245 | Schwarz criterion | -8.450364 |
| Log likelihood | 123.3034 | Hannan-Quinn criter. | -8.549464 |
| F-statistic | 18.86563 | Durbin-Watson stat | 2.161623 |
| Prob(F-statistic) | 0.000010 | | |

*S1.5.2. Unit-root test for GSENSEX*

Null Hypothesis: GSENSEX has a unit root

Exogenous: None

Lag Length: 0 (Automatic - based on SIC, maxlag=7)

| | | t-Statistic | Prob.* |
|---|---|---|---|
| Augmented Dickey-Fuller test statistic | | -6.063706 | 0.0000 |
| Test critical values: | 1% level | -2.647120 | |
| | 5% level | -1.952910 | |
| | 10% level | -1.610011 | |

*MacKinnon (1996) one-sided p-values.

Augmented Dickey-Fuller Test Equation

Dependent Variable: D(GSENSEX)

Method: Least Squares

Date: 07/08/20   Time: 12:45

Sample (adjusted): 6/02/2020 6/30/2020

Included observations: 29 after adjustments

| Variable | Coefficient | Std. Error | t-Statistic | Prob. |
|---|---|---|---|---|
| GSENSEX(-1) | -1.020402 | 0.168280 | -6.063706 | 0.0000 |

| | | | |
|---|---|---|---|
| R-squared | 0.565833 | Mean dependent var | -0.000980 |
| Adjusted R-squared | 0.565833 | S.D. dependent var | 0.015219 |
| S.E. of regression | 0.010028 | Akaike info criterion | -6.333000 |
| Sum squared resid | 0.002816 | Schwarz criterion | -6.285852 |
| Log likelihood | 92.82850 | Hannan-Quinn criter. | -6.318234 |
| Durbin-Watson stat | 2.210104 | | |

### *S1.5.3. Unit-root test for GEX*

Null Hypothesis: GEX has a unit root

Exogenous: None

Lag Length: 0 (Automatic - based on SIC, maxlag=7)

| | | t-Statistic | Prob.* |
|---|---|---|---|
| Augmented Dickey-Fuller test statistic | | -4.720960 | 0.0000 |
| Test critical values: | 1% level | -2.647120 | |
| | 5% level | -1.952910 | |
| | 10% level | -1.610011 | |

*MacKinnon (1996) one-sided p-values.

Augmented Dickey-Fuller Test Equation

Dependent Variable: D(GEX)

Method: Least Squares

Date: 07/08/20   Time: 12:46

Sample (adjusted): 6/02/2020 6/30/2020

Included observations: 29 after adjustments

| Variable | Coefficient | Std. Error | t-Statistic | Prob. |
|---|---|---|---|---|
| GEX(-1) | -0.881777 | 0.186779 | -4.720960 | 0.0001 |

| | | | |
|---|---|---|---|
| R-squared | 0.443046 | Mean dependent var | 5.06E-05 |
| Adjusted R-squared | 0.443046 | S.D. dependent var | 0.003081 |
| S.E. of regression | 0.002300 | Akaike info criterion | -9.278205 |
| Sum squared resid | 0.000148 | Schwarz criterion | -9.231057 |
| Log likelihood | 135.5340 | Hannan-Quinn criter. | -9.263439 |
| Durbin-Watson stat | 1.899165 | | |

### *S1.6. Round f: Lockdown 1.0 to Unlock 1.0 (March 25 – June 30)*

### *S1.6.1. Unit-root test for GROWTHC*

Null Hypothesis: GROWTHC has a unit root
Exogenous: None
Lag Length: 11 (Automatic - based on SIC, maxlag=11)

| | | t-Statistic | Prob.* |
|---|---|---|---|
| Augmented Dickey-Fuller test statistic | | -2.577656 | 0.0104 |
| Test critical values: | 1% level | -2.592129 | |
| | 5% level | -1.944619 | |
| | 10% level | -1.614288 | |

*MacKinnon (1996) one-sided p-values.

Augmented Dickey-Fuller Test Equation
Dependent Variable: D(GROWTHC)
Method: Least Squares

Date: 07/08/20   Time: 15:58

Sample (adjusted): 4/06/2020 6/30/2020

Included observations: 86 after adjustments

| Variable | Coefficient | Std. Error | t-Statistic | Prob. |
|---|---|---|---|---|
| GROWTHC(-1) | -0.075050 | 0.029115 | -2.577656 | 0.0119 |
| D(GROWTHC(-1)) | -0.743955 | 0.101624 | -7.320633 | 0.0000 |
| D(GROWTHC(-2)) | -0.411056 | 0.122011 | -3.369021 | 0.0012 |
| D(GROWTHC(-3)) | -0.366975 | 0.112700 | -3.256203 | 0.0017 |
| D(GROWTHC(-4)) | 0.066533 | 0.119279 | 0.557798 | 0.5787 |
| D(GROWTHC(-5)) | 0.383748 | 0.107652 | 3.564708 | 0.0006 |
| D(GROWTHC(-6)) | 0.211301 | 0.104451 | 2.022957 | 0.0467 |
| D(GROWTHC(-7)) | 0.127963 | 0.084095 | 1.521662 | 0.1324 |
| D(GROWTHC(-8)) | 0.068154 | 0.061528 | 1.107686 | 0.2716 |
| D(GROWTHC(-9)) | -0.106908 | 0.055595 | -1.922985 | 0.0583 |
| D(GROWTHC(-10)) | -0.158045 | 0.048610 | -3.251271 | 0.0017 |
| D(GROWTHC(-11)) | -0.124442 | 0.036883 | -3.373981 | 0.0012 |

| | | | |
|---|---|---|---|
| R-squared | 0.858176 | Mean dependent var | -0.001527 |
| Adjusted R-squared | 0.837094 | S.D. dependent var | 0.033244 |
| S.E. of regression | 0.013418 | Akaike info criterion | -5.655668 |
| Sum squared resid | 0.013323 | Schwarz criterion | -5.313201 |
| Log likelihood | 255.1937 | Hannan-Quinn criter. | -5.517841 |
| Durbin-Watson stat | 2.149824 | | |

*S1.6.2. Unit-root test for GSENSEX*

Null Hypothesis: GSENSEX has a unit root

Exogenous: None

Lag Length: 0 (Automatic - based on SIC, maxlag=11)

|                                        | t-Statistic | Prob.* |
|----------------------------------------|-------------|--------|
| Augmented Dickey-Fuller test statistic | -10.46675   | 0.0000 |
| Test critical values: 1% level         | -2.589020   |        |
| 5% level                               | -1.944175   |        |
| 10% level                              | -1.614554   |        |

*MacKinnon (1996) one-sided p-values.

Augmented Dickey-Fuller Test Equation

Dependent Variable: D(GSENSEX)

Method: Least Squares

Date: 07/08/20   Time: 15:59

Sample (adjusted): 3/26/2020 6/30/2020

Included observations: 97 after adjustments

| Variable    | Coefficient | Std. Error | t-Statistic | Prob.  |
|-------------|-------------|------------|-------------|--------|
| GSENSEX(-1) | -1.000680   | 0.095606   | -10.46675   | 0.0000 |

| R-squared          | 0.532641 | Mean dependent var    | -0.000733 |
|--------------------|----------|-----------------------|-----------|
| Adjusted R-squared | 0.532641 | S.D. dependent var    | 0.027881  |
| S.E. of regression | 0.019061 | Akaike info criterion | -5.072111 |
| Sum squared resid  | 0.034878 | Schwarz criterion     | -5.045568 |
| Log likelihood     | 246.9974 | Hannan-Quinn criter.  | -5.061378 |
| Durbin-Watson stat | 2.127587 |                       |           |

*S1.6.3. Unit-root test for GEX*

Null Hypothesis: GEX has a unit root

Exogenous: None

Lag Length: 0 (Automatic - based on SIC, maxlag=11)

|                                        | t-Statistic | Prob.*  |
|----------------------------------------|-------------|---------|
| Augmented Dickey-Fuller test statistic | -10.79482   | 0.0000  |
| Test critical values: 1% level         | -2.589020   |         |
| 5% level                               | -1.944175   |         |
| 10% level                              | -1.614554   |         |

*MacKinnon (1996) one-sided p-values.

Augmented Dickey-Fuller Test Equation

Dependent Variable: D(GEX)

Method: Least Squares

Date: 07/08/20   Time: 15:59

Sample (adjusted): 3/26/2020 6/30/2020

Included observations: 97 after adjustments

| Variable | Coefficient | Std. Error | t-Statistic | Prob. |
|----------|-------------|------------|-------------|-------|
| GEX(-1)  | -1.096586   | 0.101584   | -10.79482   | 0.0000 |

| | | | |
|---|---|---|---|
| R-squared | 0.548295 | Mean dependent var | 5.58E-06 |
| Adjusted R-squared | 0.548295 | S.D. dependent var | 0.006167 |
| S.E. of regression | 0.004145 | Akaike info criterion | -8.123748 |
| Sum squared resid | 0.001649 | Schwarz criterion | -8.097205 |
| Log likelihood | 395.0018 | Hannan-Quinn criter. | -8.113016 |
| Durbin-Watson stat | 1.795259 | | |

### *S1.7. Round h: Pre-lockdown to Unlock 1.0 (March 11 – June 30)*

#### *S1.7.1. Unit-root test for GROWTHC*

Null Hypothesis: D(GROWTHC) has a unit root

Exogenous: None

Lag Length: 3 (Automatic - based on SIC, maxlag=12)

|  | t-Statistic | Prob.* |
|---|---|---|
| Augmented Dickey-Fuller test statistic | -11.07113 | 0.0000 |
| Test critical values: 1% level | -2.586753 |  |
| 5% level | -1.943853 |  |
| 10% level | -1.614749 |  |

*MacKinnon (1996) one-sided p-values.

Augmented Dickey-Fuller Test Equation

Dependent Variable: D(GROWTHC,2)

Method: Least Squares

Date: 07/08/20   Time: 16:04

Sample (adjusted): 3/16/2020 6/30/2020

Included observations: 107 after adjustments

| Variable | Coefficient | Std. Error | t-Statistic | Prob. |
|---|---|---|---|---|
| D(GROWTHC(-1)) | -3.225016 | 0.291300 | -11.07113 | 0.0000 |
| D(GROWTHC(-1),2) | 1.444382 | 0.240247 | 6.012078 | 0.0000 |
| D(GROWTHC(-2),2) | 0.961579 | 0.169819 | 5.662363 | 0.0000 |
| D(GROWTHC(-3),2) | 0.436950 | 0.086341 | 5.060775 | 0.0000 |
| R-squared | 0.833401 | Mean dependent var | 0.001263 |
| Adjusted R-squared | 0.828548 | S.D. dependent var | 0.128502 |
| S.E. of regression | 0.053209 | Akaike info criterion | -2.992530 |
| Sum squared resid | 0.291608 | Schwarz criterion | -2.892611 |
| Log likelihood | 164.1004 | Hannan-Quinn criter. | -2.952024 |
| Durbin-Watson stat | 1.951236 |  |  |

*S1.7.2. Unit-root test for GSENSEX*

Null Hypothesis: GSENSEX has a unit root

Exogenous: None

Lag Length: 0 (Automatic - based on SIC, maxlag=12)

|  |  | t-Statistic | Prob.* |
|---|---|---|---|
| Augmented Dickey-Fuller test statistic |  | -10.70777 | 0.0000 |
| Test critical values: | 1% level | -2.585962 |  |
|  | 5% level | -1.943741 |  |
|  | 10% level | -1.614818 |  |

*MacKinnon (1996) one-sided p-values.

Augmented Dickey-Fuller Test Equation

Dependent Variable: D(GSENSEX)

Method: Least Squares

Date: 07/08/20   Time: 16:05

Sample (adjusted): 3/12/2020 6/30/2020

Included observations: 111 after adjustments

| Variable | Coefficient | Std. Error | t-Statistic | Prob. |
|---|---|---|---|---|
| GSENSEX(-1) | -1.020718 | 0.095325 | -10.70777 | 0.0000 |
| R-squared | 0.510363 | Mean dependent var | | -2.76E-05 |
| Adjusted R-squared | 0.510363 | S.D. dependent var | | 0.038513 |
| S.E. of regression | 0.026949 | Akaike info criterion | | -4.380768 |
| Sum squared resid | 0.079888 | Schwarz criterion | | -4.356358 |
| Log likelihood | 244.1326 | Hannan-Quinn criter. | | -4.370865 |
| Durbin-Watson stat | 1.912996 | | | |

*S1.7.3. Unit-root test for GEX*

Null Hypothesis: GEX has a unit root

Exogenous: None

Lag Length: 0 (Automatic - based on SIC, maxlag=12)

|  |  | t-Statistic | Prob.* |
|---|---|---|---|
| Augmented Dickey-Fuller test statistic |  | -11.77495 | 0.0000 |
| Test critical values: | 1% level | -2.585962 |  |
|  | 5% level | -1.943741 |  |
|  | 10% level | -1.614818 |  |

*MacKinnon (1996) one-sided p-values.

Augmented Dickey-Fuller Test Equation

Dependent Variable: D(GEX)

Method: Least Squares

Date: 07/08/20   Time: 16:05

Sample (adjusted): 3/12/2020 6/30/2020

Included observations: 111 after adjustments

| Variable | Coefficient | Std. Error | t-Statistic | Prob. |
|---|---|---|---|---|
| GEX(-1) | -1.110241 | 0.094288 | -11.77495 | 0.0000 |
| R-squared | 0.557590 | Mean dependent var | 4.40E-05 |  |
| Adjusted R-squared | 0.557590 | S.D. dependent var | 0.006579 |  |
| S.E. of regression | 0.004376 | Akaike info criterion | -8.016300 |  |
| Sum squared resid | 0.002107 | Schwarz criterion | -7.991890 |  |
| Log likelihood | 445.9047 | Hannan-Quinn criter. | -8.006398 |  |
| Durbin-Watson stat | 1.924101 |  |  |  |

## S2. Results of multiple linear regression models across different rounds

### S2.1. Round a: Lockdown 1.0 (March 25 – April 14)

#### S2.1.1. MLRM for response GROWTHC

Dependent Variable: GROWTHC

Method: Least Squares

Date: 07/08/20   Time: 16:07

Sample: 3/25/2020 4/14/2020

Included observations: 21

| Variable | Coefficient | Std. Error | t-Statistic | Prob. |
|---|---|---|---|---|
| C | 0.163634 | 0.020572 | 7.954115 | 0.0000 |
| GSENSEX | -0.176696 | 0.677014 | -0.260993 | 0.7971 |
| GEX | 6.817109 | 3.703791 | 1.840576 | 0.0822* |

| | | | |
|---|---|---|---|
| R-squared | 0.203451 | Mean dependent var | 0.160851 |
| Adjusted R-squared | 0.114945 | S.D. dependent var | 0.097658 |
| S.E. of regression | 0.091874 | Akaike info criterion | -1.805235 |
| Sum squared resid | 0.151935 | Schwarz criterion | -1.656017 |
| Log likelihood | 21.95497 | Hannan-Quinn criter. | -1.772851 |
| F-statistic | 2.298734 | Durbin-Watson stat | 1.837904 |
| Prob(F-statistic) | 0.129096 | | |

#### S2.1.2. MLRM for response GSENSEX

Dependent Variable: GSENSEX

Method: Least Squares

Date: 07/08/20   Time: 16:08

Sample: 3/25/2020 4/14/2020

Included observations: 21

| Variable | Coefficient | Std. Error | t-Statistic | Prob. |
|---|---|---|---|---|
| C | 0.010188 | 0.014999 | 0.679268 | 0.5056 |
| GROWTHC | -0.021336 | 0.081750 | -0.260993 | 0.7971 |
| GEX | -2.060027 | 1.316234 | -1.565092 | 0.1350 |
| R-squared | 0.166906 | Mean dependent var | | 0.007212 |
| Adjusted R-squared | 0.074340 | S.D. dependent var | | 0.033183 |
| S.E. of regression | 0.031926 | Akaike info criterion | | -3.919257 |
| Sum squared resid | 0.018346 | Schwarz criterion | | -3.770040 |
| Log likelihood | 44.15220 | Hannan-Quinn criter. | | -3.886873 |
| F-statistic | 1.803097 | Durbin-Watson stat | | 1.958781 |
| Prob(F-statistic) | 0.193307 | | | |

### S2.1.3. MLRM for response GEX

Dependent Variable: GEX
Method: Least Squares
Date: 07/08/20   Time: 16:26
Sample: 3/25/2020 4/14/2020
Included observations: 21

| Variable | Coefficient | Std. Error | t-Statistic | Prob. |
|---|---|---|---|---|
| C | -0.003539 | 0.002412 | -1.467526 | 0.1595 |
| GROWTHC | 0.023235 | 0.012624 | 1.840576 | 0.0822* |
| GSENSEX | -0.058147 | 0.037152 | -1.565092 | 0.1350 |
| R-squared | 0.296211 | Mean dependent var | | -0.000221 |
| Adjusted R-squared | 0.218012 | S.D. dependent var | | 0.006065 |
| S.E. of regression | 0.005364 | Akaike info criterion | | -7.486765 |
| Sum squared resid | 0.000518 | Schwarz criterion | | -7.337548 |
| Log likelihood | 81.61104 | Hannan-Quinn criter. | | -7.454381 |

| F-statistic | 3.787919 | Durbin-Watson stat | 2.467532 |
| Prob(F-statistic) | 0.042363* | | |

### S2.2. Round b: Lockdown 2.0 (April 15 – May 03)

#### S2.2.1. MLRM for response GROWTHC

Dependent Variable: GROWTHC
Method: Least Squares
Date: 07/08/20   Time: 16:39
Sample: 4/15/2020 5/03/2020
Included observations: 19

| Variable | Coefficient | Std. Error | t-Statistic | Prob. |
|---|---|---|---|---|
| C | 0.072971 | 0.004116 | 17.72677 | 0.0000 |
| GSENSEX | -0.291959 | 0.413779 | -0.705592 | 0.4906 |
| GEX | 0.100851 | 1.530551 | 0.065892 | 0.9483 |

| | | | |
|---|---|---|---|
| R-squared | 0.087252 | Mean dependent var | 0.071411 |
| Adjusted R-squared | -0.026841 | S.D. dependent var | 0.016682 |
| S.E. of regression | 0.016905 | Akaike info criterion | -5.178533 |
| Sum squared resid | 0.004572 | Schwarz criterion | -5.029411 |
| Log likelihood | 52.19607 | Hannan-Quinn criter. | -5.153296 |
| F-statistic | 0.764745 | Durbin-Watson stat | 1.741838 |
| Prob(F-statistic) | 0.481732 | | |

#### S2.2.2. MLRM for response GSENSEX

Dependent Variable: GSENSEX
Method: Least Squares
Date: 07/08/20   Time: 16:39
Sample: 4/15/2020 5/03/2020

Included observations: 19

| Variable | Coefficient | Std. Error | t-Statistic | Prob. |
|---|---|---|---|---|
| C | 0.010307 | 0.010825 | 0.952161 | 0.3552 |
| GROWTHC | -0.103361 | 0.146489 | -0.705592 | 0.4906 |
| GEX | -2.822533 | 0.575886 | -4.901204 | 0.0002 |

| | | | |
|---|---|---|---|
| R-squared | 0.635001 | Mean dependent var | 0.005080 |
| Adjusted R-squared | 0.589376 | S.D. dependent var | 0.015696 |
| S.E. of regression | 0.010058 | Akaike info criterion | -6.216918 |
| Sum squared resid | 0.001619 | Schwarz criterion | -6.067796 |
| Log likelihood | 62.06072 | Hannan-Quinn criter. | -6.191680 |
| F-statistic | 13.91786 | Durbin-Watson stat | 1.615159 |
| Prob(F-statistic) | 0.000315* | | |

*S2.2.3. MLRM for response GEX*

Dependent Variable: GEX
Method: Least Squares
Date: 07/08/20   Time: 16:40
Sample: 4/15/2020 5/03/2020
Included observations: 19

| Variable | Coefficient | Std. Error | t-Statistic | Prob. |
|---|---|---|---|---|
| C | 0.000125 | 0.003054 | 0.040914 | 0.9679 |
| GROWTHC | 0.002690 | 0.040824 | 0.065892 | 0.9483 |
| GSENSEX | -0.212652 | 0.043388 | -4.901204 | 0.0002* |

| | | | |
|---|---|---|---|
| R-squared | 0.623746 | Mean dependent var | -0.000763 |
| Adjusted R-squared | 0.576714 | S.D. dependent var | 0.004243 |
| S.E. of regression | 0.002761 | Akaike info criterion | -8.802649 |
| Sum squared resid | 0.000122 | Schwarz criterion | -8.653527 |
| Log likelihood | 86.62517 | Hannan-Quinn criter. | -8.777412 |
| F-statistic | 13.26221 | Durbin-Watson stat | 1.237533 |
| Prob(F-statistic) | 0.000402* | | |

### *S2.3. Round c: Lockdown 3.0 (May 04 – May 17)*

#### *S2.3.1. MLRM for response GROWTHC*

Dependent Variable: GROWTHC

Method: Least Squares

Date: 07/08/20   Time: 16:42

Sample: 5/04/2020 5/17/2020

Included observations: 14

| Variable | Coefficient | Std. Error | t-Statistic | Prob. |
|---|---|---|---|---|
| C | 0.057291 | 0.002811 | 20.37766 | 0.0000 |
| GSENSEX | -0.324580 | 0.157552 | -2.060143 | 0.0638* |
| GEX | 0.729374 | 0.600518 | 1.214575 | 0.2500 |

| | | | |
|---|---|---|---|
| R-squared | 0.429432 | Mean dependent var | 0.059747 |
| Adjusted R-squared | 0.325692 | S.D. dependent var | 0.012190 |
| S.E. of regression | 0.010010 | Akaike info criterion | -6.183114 |
| Sum squared resid | 0.001102 | Schwarz criterion | -6.046174 |
| Log likelihood | 46.28180 | Hannan-Quinn criter. | -6.195791 |
| F-statistic | 4.139519 | Durbin-Watson stat | 1.294806 |
| Prob(F-statistic) | 0.045676 | | |

#### *S2.3.2. MLRM for response GSENSEX*

Dependent Variable: GSENSEX

Method: Least Squares

Date: 07/08/20   Time: 16:42

Sample: 5/04/2020 5/17/2020

Included observations: 14

| Variable | Coefficient | Std. Error | t-Statistic | Prob. |
|---|---|---|---|---|
| C | 0.045904 | 0.024857 | 1.846707 | 0.0918 |
| GROWTHC | -0.857766 | 0.416362 | -2.060143 | 0.0638* |
| GEX | -0.283725 | 1.036099 | -0.273839 | 0.7893 |

| | | | |
|---|---|---|---|
| R-squared | 0.357295 | Mean dependent var | -0.005595 |
| Adjusted R-squared | 0.240440 | S.D. dependent var | 0.018671 |
| S.E. of regression | 0.016272 | Akaike info criterion | -5.211315 |
| Sum squared resid | 0.002913 | Schwarz criterion | -5.074375 |
| Log likelihood | 39.47921 | Hannan-Quinn criter. | -5.223992 |
| F-statistic | 3.057587 | Durbin-Watson stat | 2.276306 |
| Prob(F-statistic) | 0.087915 | | |

*S2.3.3. MLRM for response GEX*

Dependent Variable: GEX

Method: Least Squares

Date: 07/08/20   Time: 16:43

Sample: 5/04/2020 5/17/2020

Included observations: 14

| Variable | Coefficient | Std. Error | t-Statistic | Prob. |
|---|---|---|---|---|
| C | -0.008942 | 0.007798 | -1.146666 | 0.2759 |
| GROWTHC | 0.162125 | 0.133483 | 1.214575 | 0.2500 |
| GSENSEX | -0.023864 | 0.087148 | -0.273839 | 0.7893 |

| | | | |
|---|---|---|---|
| R-squared | 0.214641 | Mean dependent var | 0.000878 |
| Adjusted R-squared | 0.071848 | S.D. dependent var | 0.004898 |

| | | | | |
|---|---|---|---|---|
| S.E. of regression | 0.004719 | Akaike info criterion | -7.686930 | |
| Sum squared resid | 0.000245 | Schwarz criterion | -7.549989 | |
| Log likelihood | 56.80851 | Hannan-Quinn criter. | -7.699607 | |
| F-statistic | 1.503164 | Durbin-Watson stat | 2.639018 | |
| Prob(F-statistic) | 0.264775 | | | |

### *S2.4. Round d: Lockdown 4.0 (May 18 – May 31)*

#### *S2.4.1. MLRM for response GROWTHC*

Dependent Variable: GROWTHC

Method: Least Squares

Date: 07/08/20   Time: 17:00

Sample: 5/18/2020 5/31/2020

Included observations: 14

| Variable | Coefficient | Std. Error | t-Statistic | Prob. |
|---|---|---|---|---|
| C | 0.050839 | 0.001318 | 38.56308 | 0.0000 |
| GSENSEX | -0.058791 | 0.091211 | -0.644559 | 0.5324 |
| GEX | 0.964298 | 0.464557 | 2.075739 | 0.0622* |

| | | | |
|---|---|---|---|
| R-squared | 0.281463 | Mean dependent var | 0.050459 |
| Adjusted R-squared | 0.150820 | S.D. dependent var | 0.005187 |
| S.E. of regression | 0.004780 | Akaike info criterion | -7.661285 |
| Sum squared resid | 0.000251 | Schwarz criterion | -7.524344 |
| Log likelihood | 56.62900 | Hannan-Quinn criter. | -7.673961 |
| F-statistic | 2.154444 | Durbin-Watson stat | 1.198755 |
| Prob(F-statistic) | 0.162357 | | |

#### *S2.4.2. MLRM for response GSENSEX*

Dependent Variable: GSENSEX

Method: Least Squares

Date: 07/08/20   Time: 17:01

Sample: 5/18/2020 5/31/2020

Included observations: 14

| Variable | Coefficient | Std. Error | t-Statistic | Prob. |
|---|---|---|---|---|
| C | 0.034774 | 0.048810 | 0.712439 | 0.4910 |
| GROWTHC | -0.619044 | 0.960414 | -0.644559 | 0.5324 |
| GEX | 2.152433 | 1.655699 | 1.300015 | 0.2202 |

| | | | |
|---|---|---|---|
| R-squared | 0.133190 | Mean dependent var | 0.003097 |
| Adjusted R-squared | -0.024412 | S.D. dependent var | 0.015325 |
| S.E. of regression | 0.015511 | Akaike info criterion | -5.307098 |
| Sum squared resid | 0.002647 | Schwarz criterion | -5.170157 |
| Log likelihood | 40.14969 | Hannan-Quinn criter. | -5.319775 |
| F-statistic | 0.845101 | Durbin-Watson stat | 1.316734 |
| Prob(F-statistic) | 0.455599 | | |

*S2.4.3. MLRM for response GEX*

Dependent Variable: GEX

Method: Least Squares

Date: 07/08/20   Time: 17:01

Sample: 5/18/2020 5/31/2020

Included observations: 14

| Variable | Coefficient | Std. Error | t-Statistic | Prob. |
|---|---|---|---|---|
| C | -0.015124 | 0.007131 | -2.120914 | 0.0575 |
| GROWTHC | 0.291874 | 0.140612 | 2.075739 | 0.0622* |
| GSENSEX | 0.061873 | 0.047594 | 1.300015 | 0.2202 |

| | | | |
|---|---|---|---|
| R-squared | 0.353633 | Mean dependent var | -0.000205 |
| Adjusted R-squared | 0.236111 | S.D. dependent var | 0.003009 |
| S.E. of regression | 0.002630 | Akaike info criterion | -8.856362 |
| Sum squared resid | 7.61E-05 | Schwarz criterion | -8.719421 |
| Log likelihood | 64.99453 | Hannan-Quinn criter. | -8.869038 |
| F-statistic | 3.009094 | Durbin-Watson stat | 2.524826 |
| Prob(F-statistic) | 0.090706* | | |

## S2.5. Round e: Unlock 1.0 (June 01 – June 30)

### S2.5.1. MLRM for response GSENSEX

Dependent Variable: GSENSEX
Method: Least Squares
Date: 07/08/20   Time: 17:03
Sample: 6/01/2020 6/30/2020
Included observations: 30

| Variable | Coefficient | Std. Error | t-Statistic | Prob. |
|---|---|---|---|---|
| C | 0.002500 | 0.001911 | 1.308231 | 0.2014 |
| GEX | -1.392128 | 0.849926 | -1.637940 | 0.1126 |

| | | | |
|---|---|---|---|
| R-squared | 0.087438 | Mean dependent var | 0.002527 |
| Adjusted R-squared | 0.054847 | S.D. dependent var | 0.010766 |
| S.E. of regression | 0.010467 | Akaike info criterion | -6.216930 |
| Sum squared resid | 0.003067 | Schwarz criterion | -6.123517 |
| Log likelihood | 95.25395 | Hannan-Quinn criter. | -6.187046 |
| F-statistic | 2.682849 | Durbin-Watson stat | 2.126028 |
| Prob(F-statistic) | 0.112625 | | |

## S2.6. Round f: Lockdown 1.0 to Unlock 1.0 (March 25 – June 30)

*S2.6.1. MLRM for response GROWTHC*

Dependent Variable: GROWTHC

Method: Least Squares

Date: 07/08/20   Time: 17:05

Sample: 3/25/2020 6/30/2020

Included observations: 98

| Variable | Coefficient | Std. Error | t-Statistic | Prob. |
|---|---|---|---|---|
| C | 0.075868 | 0.006517 | 11.64243 | 0.0000 |
| GSENSEX | 0.079016 | 0.351351 | 0.224892 | 0.8225 |
| GEX | 3.556297 | 1.698950 | 2.093233 | 0.0390* |
| R-squared | 0.048075 | Mean dependent var | 0.075727 |
| Adjusted R-squared | 0.028034 | S.D. dependent var | 0.064690 |
| S.E. of regression | 0.063777 | Akaike info criterion | -2.636708 |
| Sum squared resid | 0.386416 | Schwarz criterion | -2.557576 |
| Log likelihood | 132.1987 | Hannan-Quinn criter. | -2.604701 |
| F-statistic | 2.398873 | Durbin-Watson stat | 0.935450 |
| Prob(F-statistic) | 0.096302* | | |

*S2.6.2. MLRM for response GSENSEX*

Dependent Variable: GSENSEX

Method: Least Squares

Date: 07/08/20   Time: 17:06

Sample: 3/25/2020 6/30/2020

Included observations: 98

| Variable | Coefficient | Std. Error | t-Statistic | Prob. |
|---|---|---|---|---|
| C | 0.002236 | 0.002955 | 0.756779 | 0.4511 |
| GROWTHC | 0.006734 | 0.029944 | 0.224892 | 0.8225 |
| GEX | -1.913943 | 0.467740 | -4.091900 | 0.0001* |

| | | | |
|---|---|---|---|
| R-squared | 0.153385 | Mean dependent var | 0.002947 |
| Adjusted R-squared | 0.135561 | S.D. dependent var | 0.020025 |
| S.E. of regression | 0.018619 | Akaike info criterion | -5.099176 |
| Sum squared resid | 0.032932 | Schwarz criterion | -5.020044 |
| Log likelihood | 252.8596 | Hannan-Quinn criter. | -5.067169 |
| F-statistic | 8.605763 | Durbin-Watson stat | 2.130143 |
| Prob(F-statistic) | 0.000367* | | |

*S2.6.3. MLRM for response GEX*

Dependent Variable: GEX

Method: Least Squares

Date: 07/08/20   Time: 17:06

Sample: 3/25/2020 6/30/2020

Included observations: 98

| Variable | Coefficient | Std. Error | t-Statistic | Prob. |
|---|---|---|---|---|
| C | -0.000813 | 0.000594 | -1.370114 | 0.1739 |
| GROWTHC | 0.012397 | 0.005923 | 2.093233 | 0.0390* |
| GSENSEX | -0.078289 | 0.019133 | -4.091900 | 0.0001* |

| | | | |
|---|---|---|---|
| R-squared | 0.190280 | Mean dependent var | -0.000105 |
| Adjusted R-squared | 0.173233 | S.D. dependent var | 0.004141 |
| S.E. of regression | 0.003766 | Akaike info criterion | -8.295696 |
| Sum squared resid | 0.001347 | Schwarz criterion | -8.216564 |
| Log likelihood | 409.4891 | Hannan-Quinn criter. | -8.263689 |
| F-statistic | 11.16226 | Durbin-Watson stat | 2.357087 |
| Prob(F-statistic) | 0.000044* | | |

## S3. Correlation summaries across different rounds

### S3.1. Round a: Lockdown 1.0 (March 25 – April 14)

|          | GROWTHC              | GSENSEX              | GEX                  |
|----------|----------------------|----------------------|----------------------|
| GROWTHC  | 1                    | -0.2313756707880113  | 0.4477010580872346   |
| GSENSEX  | -0.2313756707880113  | 1                    | -0.4046639177145718  |
| GEX      | 0.4477010580872346   | -0.4046639177145718  | 1                    |

### S3.2. Round b: Lockdown 2.0 (April 15 – May 03)

|          | GROWTHC              | GSENSEX              | GEX                  |
|----------|----------------------|----------------------|----------------------|
| GROWTHC  | 1                    | -0.2949655590059144  | 0.2425924212055322   |
| GSENSEX  | -0.2949655590059144  | 1                    | -0.7897109760059276  |
| GEX      | 0.2425924212055322   | -0.7897109760059276  | 1                    |

### S3.3. Round c: Lockdown 3.0 (May 04 – May 17)

|          | GROWTHC              | GSENSEX              | GEX                  |
|----------|----------------------|----------------------|----------------------|
| GROWTHC  | 1                    | -0.594065752176441   | 0.4574787175868636   |
| GSENSEX  | -0.594065752176441   | 1                    | -0.3306317126371629  |
| GEX      | 0.4574787175868636   | -0.3306317126371629  | 1                    |

### S3.4. Round d: Lockdown 4.0 (May 18 – May 31)

|          | GROWTHC              | GSENSEX              | GEX                  |
|----------|----------------------|----------------------|----------------------|
| GROWTHC  | 1                    | 0.003590958809302893 | 0.5043063458499434   |
| GSENSEX  | 0.003590958809302893 | 1                    | 0.3169402889983672   |
| GEX      | 0.5043063458499434   | 0.3169402889983672   | 1                    |

### *S3.5. Round e: Unlock 1.0 (June 01 – June 30)*

|         | GROWTHC            | GSENSEX            | GEX                |
|---------|--------------------|--------------------|--------------------|
| GROWTHC | 1                  | 0.2009233369750571 | 0.1263611033248855 |
| GSENSEX | 0.2009233369750571 | 1                  | -0.2956992777823893|
| GEX     | 0.1263611033248855 | -0.2956992777823893| 1                  |

### *S3.6. Round f: Lockdown 1.0 to Unlock 1.0 (March 25 – June 30)*

|         | GROWTHC             | GSENSEX             | GEX                 |
|---------|---------------------|---------------------|---------------------|
| GROWTHC | 1                   | -0.064572941470039  | 0.2181006766284255  |
| GSENSEX | -0.064572941470039  | 1                   | -0.3910676790329568 |
| GEX     | 0.2181006766284255  | -0.3910676790329568 | 1                   |

### *S3.7. Round g: Pre-lockdown to Lockdown 1.0 (March 11 – April 14)*

|         | GROWTHC             | GSENSEX             | GEX                 |
|---------|---------------------|---------------------|---------------------|
| GROWTHC | 1                   | -0.1284512832334094 | 0.4014368887815389  |
| GSENSEX | -0.1284512832334094 | 1                   | -0.5201629315803736 |
| GEX     | 0.4014368887815389  | -0.5201629315803736 | 1                   |

### *S3.8. Round h: Pre-lockdown to Unlock 1.0 (March 11 – June 30)*

|         | GROWTHC             | GSENSEX             | GEX                 |
|---------|---------------------|---------------------|---------------------|
| GROWTHC | 1                   | -0.1531029244245676 | 0.2808517762927862  |
| GSENSEX | -0.1531029244245676 | 1                   | -0.4680084500326706 |
| GEX     | 0.2808517762927862  | -0.4680084500326706 | 1                   |

### *S4. Summaries of vector autoregression models across different rounds*

### *S4.1. Round a: Lockdown 1.0 (March 25 – April 14)*

Vector Autoregression Estimates

Date: 07/10/20   Time: 18:06

Sample (adjusted): 3/26/2020 4/14/2020

Included observations: 20 after adjustments

Standard errors in ( ) & t-statistics in [ ]

|  | GROWTHC | GSENSEX | GEX |
|---|---|---|---|
| GROWTHC(-1) | -0.096864 | 0.165784 | -0.017008 |
|  | (0.28060) | (0.07670) | (0.01665) |
|  | [-0.34520] | [ 2.16137] | [-1.02145] |
|  |  |  |  |
| GSENSEX(-1) | 0.088108 | -0.031461 | -0.027988 |
|  | (0.79413) | (0.21708) | (0.04712) |
|  | [ 0.11095] | [-0.14493] | [-0.59396] |
|  |  |  |  |
| GEX(-1) | 1.105948 | -1.213721 | -0.145464 |
|  | (4.74500) | (1.29705) | (0.28156) |
|  | [ 0.23308] | [-0.93576] | [-0.51664] |
|  |  |  |  |
| C | 0.173128 | -0.023198 | 0.002741 |
|  | (0.05305) | (0.01450) | (0.00315) |
|  | [ 3.26364] | [-1.59977] | [ 0.87093] |
| R-squared | 0.008675 | 0.229388 | 0.116142 |
| Adj. R-squared | -0.177198 | 0.084898 | -0.049581 |
| Sum sq. resids | 0.184703 | 0.013801 | 0.000650 |
| S.E. equation | 0.107443 | 0.029370 | 0.006375 |
| F-statistic | 0.046673 | 1.587573 | 0.700820 |
| Log likelihood | 18.46862 | 44.40862 | 74.95878 |
| Akaike AIC | -1.446862 | -4.040862 | -7.095878 |
| Schwarz SC | -1.247715 | -3.841716 | -6.896731 |
| Mean dependent | 0.157606 | 0.004083 | -0.000218 |
| S.D. dependent | 0.099027 | 0.030702 | 0.006223 |

| | |
|---|---|
| Determinant resid covariance (dof adj.) | 2.54E-10 |
| Determinant resid covariance | 1.30E-10 |
| Log likelihood | 142.4779 |
| Akaike information criterion | -13.04779 |
| Schwarz criterion | -12.45035 |
| Number of coefficients | 12 |

### S4.2. Round b: Lockdown 2.0 (April 15 – May 03)

Vector Autoregression Estimates

Date: 07/11/20   Time: 06:50

Sample (adjusted): 4/16/2020 5/03/2020

Included observations: 18 after adjustments

Standard errors in ( ) & t-statistics in [ ]

| | GROWTHC | GSENSEX | GEX |
|---|---|---|---|
| GROWTHC(-1) | 0.108998 | 0.071118 | 0.024792 |
| | (0.26222) | (0.25508) | (0.06073) |
| | [ 0.41567] | [ 0.27880] | [ 0.40821] |
| | | | |
| GSENSEX(-1) | 0.532489 | -0.038780 | 0.054235 |
| | (0.44174) | (0.42971) | (0.10231) |
| | [ 1.20543] | [-0.09025] | [ 0.53009] |
| | | | |
| GEX(-1) | 2.109313 | -0.096562 | 0.302748 |
| | (1.60531) | (1.56160) | (0.37181) |
| | [ 1.31396] | [-0.06184] | [ 0.81426] |
| C | 0.062396 | 0.000973 | -0.003043 |
| | (0.01968) | (0.01915) | (0.00456) |
| | [ 3.17020] | [ 0.05082] | [-0.66745] |
| R-squared | 0.122113 | 0.007481 | 0.062185 |
| Adj. R-squared | -0.066005 | -0.205202 | -0.138775 |
| Sum sq. resids | 0.004396 | 0.004160 | 0.000236 |
| S.E. equation | 0.017720 | 0.017238 | 0.004104 |
| F-statistic | 0.649130 | 0.035173 | 0.309440 |
| Log likelihood | 49.31588 | 49.81285 | 75.64446 |
| Akaike AIC | -5.035098 | -5.090317 | -7.960496 |
| Schwarz SC | -4.837238 | -4.892456 | -7.762636 |
| Mean dependent | 0.071340 | 0.005924 | -0.001224 |
| S.D. dependent | 0.017163 | 0.015702 | 0.003846 |

| | |
|---|---|
| Determinant resid covariance (dof adj.) | 4.34E-13 |
| Determinant resid covariance | 2.04E-13 |
| Log likelihood | 186.3632 |
| Akaike information criterion | -19.37369 |
| Schwarz criterion | -18.78011 |
| Number of coefficients | 12 |

### *S4.3. Round c: Lockdown 3.0 (May 04 – May 17)*

Vector Autoregression Estimates

Date: 07/11/20   Time: 06:53

Sample (adjusted): 5/05/2020 5/17/2020

Included observations: 13 after adjustments

Standard errors in ( ) & t-statistics in [ ]

|  | GROWTHC | GSENSEX | GEX |
|---|---|---|---|
| GROWTHC(-1) | 0.320293 | -0.030908 | -0.011748 |
|  | (0.24960) | (0.20481) | (0.14084) |
|  | [ 1.28322] | [-0.15090] | [-0.08341] |
| GSENSEX(-1) | 0.065734 | -0.267138 | -0.068037 |
|  | (0.15345) | (0.12592) | (0.08659) |
|  | [ 0.42837] | [-2.12151] | [-0.78576] |
| GEX(-1) | 0.083271 | -1.834478 | -0.273525 |
|  | (0.52877) | (0.43389) | (0.29836) |
|  | [ 0.15748] | [-4.22794] | [-0.91675] |
| C | 0.038309 | 0.000524 | 0.000744 |
|  | (0.01454) | (0.01193) | (0.00820) |
|  | [ 2.63527] | [ 0.04397] | [ 0.09069] |
| R-squared | 0.205515 | 0.706862 | 0.130332 |
| Adj. R-squared | -0.059313 | 0.609149 | -0.159558 |
| Sum sq. resids | 0.000616 | 0.000415 | 0.000196 |
| S.E. equation | 0.008276 | 0.006791 | 0.004670 |
| F-statistic | 0.776031 | 7.234076 | 0.449591 |
| Log likelihood | 46.27093 | 48.84176 | 53.71006 |
| Akaike AIC | -6.503221 | -6.898732 | -7.647702 |
| Schwarz SC | -6.329390 | -6.724901 | -7.473871 |
| Mean dependent | 0.057227 | -0.001457 | 0.000190 |
| S.D. dependent | 0.008041 | 0.010863 | 0.004337 |

| | | |
|---|---|---|
| Determinant resid covariance (dof adj.) | 4.22E-14 | |
| Determinant resid covariance | 1.40E-14 | |
| Log likelihood | 152.0070 | |

| Akaike information criterion | -21.53954 |
|---|---|
| Schwarz criterion | -21.01805 |
| Number of coefficients | 12 |

### S4.4. Round d: Lockdown 4.0 (May 18 – May 31)

Vector Autoregression Estimates

Date: 07/11/20   Time: 10:55

Sample (adjusted): 5/19/2020 5/31/2020

Included observations: 13 after adjustments

Standard errors in ( ) & t-statistics in [ ]

|  | GROWTHC | GSENSEX | GEX |
|---|---|---|---|
| GROWTHC(-1) | 0.448112 | -0.452878 | 0.039041 |
|  | (0.32142) | (0.77135) | (0.20524) |
|  | [ 1.39415] | [-0.58713] | [ 0.19022] |
| GSENSEX(-1) | -0.128920 | 0.169235 | -0.013566 |
|  | (0.09816) | (0.23555) | (0.06268) |
|  | [-1.31342] | [ 0.71846] | [-0.21644] |
| GEX(-1) | -0.301169 | -0.684895 | -0.401475 |
|  | (0.58049) | (1.39304) | (0.37066) |
|  | [-0.51882] | [-0.49165] | [-1.08313] |
| C | 0.028297 | 0.028191 | -0.002041 |
|  | (0.01647) | (0.03953) | (0.01052) |
|  | [ 1.71796] | [ 0.71320] | [-0.19401] |
| R-squared | 0.341703 | 0.151239 | 0.169498 |
| Adj. R-squared | 0.122271 | -0.131682 | -0.107336 |
| Sum sq. resids | 0.000227 | 0.001309 | 9.26E-05 |
| S.E. equation | 0.005025 | 0.012058 | 0.003208 |
| F-statistic | 1.557215 | 0.534563 | 0.612274 |
| Log likelihood | 52.75850 | 41.37858 | 58.58994 |
| Akaike AIC | -7.501307 | -5.750550 | -8.398453 |
| Schwarz SC | -7.327477 | -5.576720 | -8.224622 |
| Mean dependent | 0.050619 | 0.005979 | -2.08E-05 |
| S.D. dependent | 0.005363 | 0.011335 | 0.003049 |

| | |
|---|---|
| Determinant resid covariance (dof adj.) | 2.51E-14 |
| Determinant resid covariance | 8.33E-15 |
| Log likelihood | 155.3880 |
| Akaike information criterion | -22.05969 |
| Schwarz criterion | -21.53820 |
| Number of coefficients | 12 |

### S4.5. Round e: Unlock 1.0 (June 01 – June 30)

Vector Autoregression Estimates

Date: 07/11/20   Time: 10:59

Sample (adjusted): 6/03/2020 6/30/2020

Included observations: 28 after adjustments

Standard errors in ( ) & t-statistics in [ ]

| | D(GROWTHC) | GSENSEX | GEX |
|---|---|---|---|
| D(GROWTHC(-1)) | -0.188619 | 1.230814 | 0.095264 |
| | (0.18958) | (0.56261) | (0.13823) |
| | [-0.99494] | [ 2.18767] | [ 0.68918] |
| | | | |
| GSENSEX(-1) | 0.085598 | -0.289173 | -0.006063 |
|                | (0.06150) | (0.18252) | (0.04484) |
|                | [ 1.39180] | [-1.58435] | [-0.13521] |
|                |         |         |         |
| GEX(-1)        | -0.023454 | -0.865482 | 0.052088 |
|                | (0.26410) | (0.78378) | (0.19257) |
|                | [-0.08880] | [-1.10424] | [ 0.27049] |
|                |         |         |         |
| C              | -0.000613 | 0.002012 | 0.000239 |
|                | (0.00059) | (0.00176) | (0.00043) |
|                | [-1.03575] | [ 1.14554] | [ 0.55476] |
| R-squared      | 0.113041 | 0.218625 | 0.026743 |
| Adj. R-squared | 0.002171 | 0.120953 | -0.094914 |
| Sum sq. resids | 0.000225 | 0.001979 | 0.000119 |
| S.E. equation  | 0.003060 | 0.009080 | 0.002231 |
| F-statistic    | 1.019584 | 2.238361 | 0.219821 |
| Log likelihood | 124.5317 | 94.07367 | 133.3770 |
| Akaike AIC     | -8.609405 | -6.433833 | -9.241217 |
| Schwarz SC     | -8.419090 | -6.243518 | -9.050902 |
| Mean dependent | -0.000414 | 0.001179 | 0.000206 |
| S.D. dependent | 0.003063 | 0.009685 | 0.002132 |
| Determinant resid covariance (dof adj.) | | 2.86E-15 | |
| Determinant resid covariance | | 1.80E-15 | |
| Log likelihood | | 356.0975 | |
| Akaike information criterion | | -24.57839 | |
| Schwarz criterion | | -24.00745 | |
| Number of coefficients | | 12 | |

### *S4.6. Round f: Lockdown 1.0 to Unlock 1.0 (March 25 – June 30)*

Vector Autoregression Estimates

Date: 07/11/20   Time: 11:05

Sample (adjusted): 3/30/2020 6/30/2020

Included observations: 93 after adjustments

Standard errors in ( ) & t-statistics in [ ]

|  | GROWTHC | GSENSEX | GEX |
|---|---|---|---|
| GROWTHC(-1) | 0.106327 | 0.093560 | -0.010703 |
|  | (0.09433) | (0.04075) | (0.00934) |
|  | [ 1.12719] | [ 2.29600] | [-1.14606] |
| GROWTHC(-2) | 0.125962 | -0.158900 | 0.014734 |
|  | (0.09668) | (0.04176) | (0.00957) |
|  | [ 1.30288] | [-3.80467] | [ 1.53936] |
| GROWTHC(-3) | 0.048033 | 0.113527 | 0.002565 |
|  | (0.09708) | (0.04194) | (0.00961) |
|  | [ 0.49477] | [ 2.70703] | [ 0.26688] |
| GROWTHC(-4) | 0.079726 | -0.091596 | -0.005485 |
|  | (0.10133) | (0.04377) | (0.01003) |
|  | [ 0.78679] | [-2.09249] | [-0.54675] |
| GROWTHC(-5) | 0.547137 | 0.069865 | 0.005569 |
|  | (0.09649) | (0.04168) | (0.00955) |
|  | [ 5.67021] | [ 1.67605] | [ 0.58300] |
| GSENSEX(-1) | 0.290082 | 0.072721 | -0.005520 |
|  | (0.27116) | (0.11714) | (0.02684) |
|  | [ 1.06980] | [ 0.62082] | [-0.20561] |
| GSENSEX(-2) | -0.574185 | 0.037102 | -0.025701 |
| | (0.26446) | (0.11424) | (0.02618) |
| | [-2.17117] | [ 0.32476] | [-0.98164] |
| | | | |
| GSENSEX(-3) | -0.587609 | 0.130703 | -0.020135 |
| | (0.25823) | (0.11155) | (0.02557) |
| | [-2.27554] | [ 1.17168] | [-0.78758] |
| | | | |
| GSENSEX(-4) | 0.212495 | -0.255499 | 0.002964 |
| | (0.26171) | (0.11306) | (0.02591) |
| | [ 0.81194] | [-2.25991] | [ 0.11439] |
| | | | |
| GSENSEX(-5) | 0.219097 | -0.180139 | 0.005189 |
| | (0.23543) | (0.10170) | (0.02331) |
| | [ 0.93063] | [-1.77124] | [ 0.22264] |
| | | | |
| GEX(-1) | 2.132429 | -1.240362 | -0.011273 |
| | (1.25553) | (0.54237) | (0.12430) |
| | [ 1.69844] | [-2.28691] | [-0.09070] |
| | | | |
| GEX(-2) | -2.875951 | 1.054550 | -0.219043 |
| | (1.28579) | (0.55545) | (0.12730) |
| | [-2.23672] | [ 1.89856] | [-1.72074] |
| | | | |
| GEX(-3) | -1.244016 | -0.692982 | -0.253952 |
| | (1.32514) | (0.57245) | (0.13119) |
| | [-0.93878] | [-1.21056] | [-1.93574] |
| | | | |
| GEX(-4) | 0.724500 | -0.064064 | -0.015019 |
| | (1.23977) | (0.53557) | (0.12274) |
| | [ 0.58438] | [-0.11962] | [-0.12236] |
| GEX(-5) | 1.058369 | -0.628109 | -0.051972 |
| | (1.16425) | (0.50294) | (0.11526) |
| | [ 0.90906] | [-1.24887] | [-0.45090] |
| | | | |
| C | 0.003972 | 0.000420 | -0.000394 |
| | (0.00754) | (0.00326) | (0.00075) |
| | [ 0.52669] | [ 0.12878] | [-0.52713] |
| R-squared | 0.686938 | 0.348797 | 0.105779 |
| Adj. R-squared | 0.625952 | 0.221939 | -0.068420 |
| Sum sq. resids | 0.111975 | 0.020896 | 0.001098 |
| S.E. equation | 0.038134 | 0.016474 | 0.003775 |
| F-statistic | 11.26386 | 2.749513 | 0.607230 |
| Log likelihood | 180.6156 | 258.6754 | 395.6888 |
| Akaike AIC | -3.540120 | -5.218825 | -8.165350 |
| Schwarz SC | -3.104404 | -4.783109 | -7.729634 |
| Mean dependent | 0.072244 | 0.001871 | 1.56E-05 |
| S.D. dependent | 0.062352 | 0.018676 | 0.003652 |

| | |
|---|---:|
| Determinant resid covariance (dof adj.) | 4.14E-12 |
| Determinant resid covariance | 2.35E-12 |
| Log likelihood | 849.2679 |
| Akaike information criterion | -17.23157 |
| Schwarz criterion | -15.92442 |
| Number of coefficients | 48 |

VAR Residual Serial Correlation LM Tests

Date: 07/11/20   Time: 11:06

Sample: 3/25/2020 6/30/2020

Included observations: 93

Null hypothesis: No serial correlation at lag h

| Lag | LRE* stat | df | Prob. | Rao F-stat | df | Prob. |
|---|---|---|---|---|---|---|
| 1 | 24.33359 | 9 | 0.0038 | 2.839592 | (9, 175.4) | 0.0038 |
| 2 | 12.85928 | 9 | 0.1691 | 1.452433 | (9, 175.4) | 0.1692 |
| 3 | 12.91615 | 9 | 0.1664 | 1.459091 | (9, 175.4) | 0.1666 |
| 4 | 20.56982 | 9 | 0.0147 | 2.374738 | (9, 175.4) | 0.0147 |
| 5 | 6.490126 | 9 | 0.6900 | 0.719998 | (9, 175.4) | 0.6901 |
| 6 | 8.963914 | 9 | 0.4406 | 1.001384 | (9, 175.4) | 0.4408 |

Null hypothesis: No serial correlation at lags 1 to h

| Lag | LRE* stat | df | Prob. | Rao F-stat | df | Prob. |
|---|---|---|---|---|---|---|
| 1 | 24.33359 | 9 | 0.0038 | 2.839592 | (9, 175.4) | 0.0038 |
| 2 | 34.00352 | 18 | 0.0126 | 1.975189 | (18, 195.6) | 0.0127 |
| 3 | 45.77812 | 27 | 0.0134 | 1.783477 | (27, 193.4) | 0.0137 |
| 4 | 58.60878 | 36 | 0.0100 | 1.728888 | (36, 186.9) | 0.0105 |
| 5 | 77.10891 | 45 | 0.0020 | 1.865565 | (45, 179.0) | 0.0022 |
| 6 | 108.0107 | 54 | 0.0000 | 2.313120 | (54, 170.7) | 0.0000 |

*Edgeworth expansion corrected likelihood ratio statistic.

## S4.7. Round g: Pre-lockdown to Lockdown 1.0 (March 11 – April 14)

Vector Autoregression Estimates

Date: 07/11/20   Time: 11:24

Sample (adjusted): 3/12/2020 4/14/2020

Included observations: 34 after adjustments

Standard errors in ( ) & t-statistics in [ ]

|  | GROWTHC | GSENSEX | GEX |
|---|---|---|---|
| GROWTHC(-1) | -0.059037 | 0.109057 | -0.007271 |
|  | (0.19580) | (0.09091) | (0.01243) |
|  | [-0.30152] | [ 1.19964] | [-0.58475] |
|  |  |  |  |
| GSENSEX(-1) | 0.578975 | 0.014340 | -0.016787 |
|  | (0.44759) | (0.20781) | (0.02842) |
|  | [ 1.29354] | [ 0.06901] | [-0.59059] |
|  |  |  |  |
| GEX(-1) | 2.687893 | 0.382948 | -0.224175 |
|  | (3.50390) | (1.62685) | (0.22252) |
|  | [ 0.76712] | [ 0.23539] | [-1.00745] |
|  |  |  |  |
| C | 0.179608 | -0.022226 | 0.002212 |
|  | (0.03635) | (0.01688) | (0.00231) |
|  | [ 4.94171] | [-1.31707] | [ 0.95820] |
| R-squared | 0.053721 | 0.065679 | 0.067779 |
| Adj. R-squared | -0.040907 | -0.027753 | -0.025443 |
| Sum sq. resids | 0.272500 | 0.058743 | 0.001099 |
| S.E. equation | 0.095306 | 0.044250 | 0.006052 |
| F-statistic | 0.567705 | 0.702956 | 0.727067 |
| Log likelihood | 33.80622 | 59.89208 | 127.5317 |
| Akaike AIC | -1.753307 | -3.287770 | -7.266570 |
| Schwarz SC | -1.573735 | -3.108198 | -7.086998 |
| Mean dependent | 0.169509 | -0.003490 | 0.000876 |
| S.D. dependent | 0.093415 | 0.043649 | 0.005977 |

| | |
|---|---|
| Determinant resid covariance (dof adj.) | 3.88E-10 |
| Determinant resid covariance | 2.66E-10 |
| Log likelihood | 230.0444 |
| Akaike information criterion | -12.82614 |
| Schwarz criterion | -12.28743 |
| Number of coefficients | 12 |

### *S4.8. Round h: Pre-lockdown to Unlock 1.0 (March 11 – June 30)*

Vector Autoregression Estimates

Date: 07/11/20   Time: 11:27

Sample (adjusted): 3/16/2020 6/30/2020

Included observations: 107 after adjustments

Standard errors in ( ) & t-statistics in [ ]

| | D(GROWTHC) | GSENSEX | GEX |
|---|---|---|---|
| D(GROWTHC(-1)) | -0.809543 | 0.065822 | -0.006828 |
| | (0.09653) | (0.04221) | (0.00759) |
| | [-8.38671] | [ 1.55946] | [-0.89974] |
| | | | |
| D(GROWTHC(-2)) | -0.476650 | -0.140864 | 0.014838 |
| | (0.11388) | (0.04980) | (0.00895) |
| | [-4.18549] | [-2.82879] | [ 1.65736] |
| | | | |
| D(GROWTHC(-3)) | -0.524470 | -0.002937 | 0.011770 |
|  | (0.11008) | (0.04813) | (0.00865) |
|  | [-4.76437] | [-0.06102] | [ 1.36005] |

| D(GROWTHC(-4)) | -0.379081 | -0.108689 | 0.005617 |
|---|---|---|---|
|  | (0.09964) | (0.04357) | (0.00783) |
|  | [-3.80451] | [-2.49463] | [ 0.71701] |

| GSENSEX(-1) | 0.075213 | 0.248085 | -0.042861 |
|---|---|---|---|
|  | (0.24439) | (0.10686) | (0.01921) |
|  | [ 0.30776] | [ 2.32152] | [-2.23080] |

| GSENSEX(-2) | -0.342825 | -0.142794 | -0.026046 |
|---|---|---|---|
|  | (0.24158) | (0.10563) | (0.01899) |
|  | [-1.41912] | [-1.35179] | [-1.37141] |

| GSENSEX(-3) | -0.483460 | -0.090150 | 0.015680 |
|---|---|---|---|
|  | (0.24169) | (0.10568) | (0.01900) |
|  | [-2.00037] | [-0.85304] | [ 0.82523] |

| GSENSEX(-4) | -0.373565 | -0.026276 | -0.024239 |
|---|---|---|---|
|  | (0.23232) | (0.10159) | (0.01826) |
|  | [-1.60795] | [-0.25865] | [-1.32712] |

| GEX(-1) | 0.698684 | -0.543759 | -0.104594 |
|---|---|---|---|
|  | (1.45825) | (0.63764) | (0.11464) |
|  | [ 0.47912] | [-0.85276] | [-0.91234] |

| GEX(-2) | -3.011890 | 1.663799 | -0.215196 |
|---|---|---|---|
|  | (1.42298) | (0.62222) | (0.11187) |
|  | [-2.11661] | [ 2.67397] | [-1.92362] |

| GEX(-3) | 0.325406 | -1.465898 | -0.090598 |
|---|---|---|---|
|  | (1.40952) | (0.61634) | (0.11081) |
|  | [ 0.23086] | [-2.37841] | [-0.81758] |
| GEX(-4) | -2.055098 | 0.490872 | -0.026469 |
|  | (1.43912) | (0.62928) | (0.11314) |
|  | [-1.42802] | [ 0.78006] | [-0.23395] |
| C | -0.001863 | 0.000373 | 0.000363 |
|  | (0.00510) | (0.00223) | (0.00040) |
|  | [-0.36548] | [ 0.16734] | [ 0.90441] |
| R-squared | 0.553705 | 0.332469 | 0.202085 |
| Adj. R-squared | 0.496731 | 0.247252 | 0.100223 |
| Sum sq. resids | 0.249854 | 0.047772 | 0.001544 |
| S.E. equation | 0.051556 | 0.022544 | 0.004053 |
| F-statistic | 9.718587 | 3.901454 | 1.983915 |
| Log likelihood | 172.3680 | 260.8798 | 444.4869 |
| Akaike AIC | -2.978840 | -4.633268 | -8.065175 |
| Schwarz SC | -2.654104 | -4.308532 | -7.740439 |
| Mean dependent | -0.000701 | 0.000561 | 0.000222 |
| S.D. dependent | 0.072674 | 0.025984 | 0.004273 |
| Determinant resid covariance (dof adj.) |  | 1.56E-11 |  |
| Determinant resid covariance |  | 1.06E-11 |  |
| Log likelihood |  | 896.4413 |  |
| Akaike information criterion |  | -16.02694 |  |
| Schwarz criterion |  | -15.05273 |  |
| Number of coefficients |  | 39 |  |

## S5. Robustness tests for fitted VAR models

VAR Residual Serial Correlation LM Tests

| Null hypothesis: No serial correlation at lag h | | | | |
|---|---|---|---|---|
| Lag | LRE* stat | Prob. | Rao F-stat | Prob. |
| 1 | 24.33 | 0.00 | 2.84 | 0.00 |
| 2 | 12.86 | 0.17 | 1.45 | 0.17 |
| 3 | 12.92 | 0.17 | 1.46 | 0.17 |
| 4 | 20.57 | 0.01 | 2.37 | 0.01 |
| 5 | 6.49 | 0.69 | 0.72 | 0.69 |
| 6 | 8.96 | 0.44 | 1.00 | 0.44 |

| Null hypothesis: No serial correlation at lags 1 to h | | | | |
|---|---|---|---|---|
| Lag | LRE* stat | Prob. | Rao F-stat | Prob. |
| 1 | 24.33 | 0.00 | 2.84 | 0.00 |
| 2 | 34.00 | 0.01 | 1.98 | 0.01 |
| 3 | 45.78 | 0.01 | 1.78 | 0.01 |
| 4 | 58.61 | 0.01 | 1.73 | 0.01 |
| 5 | 77.11 | 0.00 | 1.87 | 0.00 |
| 6 | 108.01 | 0.00 | 2.31 | 0.00 |

*Edgeworth expansion corrected likelihood ratio statistic.

| VAR Lag Order Selection Criteria | | | | | | |
|---|---|---|---|---|---|---|
| Endogenous variables: GROWTHC GSENSEX GEX | | | | | | |
| Exogenous variables: C | | | | | | |
| Lag | LogL | LR | FPE | AIC | SC | HQ |
| 0 | 773.6018 | NA | 7.34E-12 | -17.1245 | -17.0412 | -17.0909 |
| 1 | 797.835 | 46.31242 | 5.23E-12 | -17.463 | -17.12969* | -17.3286 |
| 2 | 808.25 | 19.20985 | 5.07E-12 | -17.4944 | -16.9112 | -17.2592 |
| 3 | 823.083 | 26.36989 | 4.46E-12 | -17.6241 | -16.7908 | -17.288 |
| 4 | 846.8635 | 40.69104 | 3.23E-12 | -17.9525 | -16.8693 | -17.5157 |
| 5 | 872.0494 | 41.41685* | 2.26e-12* | -18.31221* | -16.979 | -17.77457* |
| 6 | 879.3946 | 11.58909 | 2.37E-12 | -18.2754 | -16.6922 | -17.637 |
| 7 | 884.8081 | 8.180359 | 2.59E-12 | -18.1957 | -16.3625 | -17.4565 |
| 8 | 895.1089 | 14.8789 | 2.55E-12 | -18.2246 | -16.1415 | -17.3846 |

* indicates lag order selected by the criterion
LR: sequential modified LR test statistic (each test at 5% level)
FPE: Final prediction error
AIC: Akaike information criterion
SC: Schwarz information criterion
HQ: Hannan-Quinn information criterion